\documentclass[journal,draftclsnofoot,onecolumn,12pt]{IEEEtran}

\usepackage{amsthm,amssymb,graphicx,multirow,amsmath,color,amsfonts}
\usepackage[update,prepend]{epstopdf}
\usepackage[noadjust]{cite}
\usepackage[latin1]{inputenc}
\usepackage{tikz}
\usetikzlibrary{arrows,calc}		
\usepackage{bbm} 
\usepackage{pdfpages}
\usepackage{tabulary}
\usepackage{multirow}
\usepackage{comment}
\usepackage{dsfont}
\usepackage{pgf,tikz}
\usepackage{mathrsfs}
\usetikzlibrary{arrows}
\usepackage{subcaption}
\usepackage{bigints}
\usepackage{mathtools}
\newcommand\numberthis{\addtocounter{equation}{1}\tag{\theequation}}
\usepackage{mathtools}
\usepackage{dsfont}
\usepackage{color}

\newcommand{\ppmf}{\mathtt{Pos_{PMF}}}
\newcommand{\pcmf}{\mathtt{Pos_{CMF}}}
\newcommand{\pmf}{\mathtt{PMF}}

\newcommand{\fir}{\Phi_{\tt r}}
\newcommand{\fiu}{\Phi_{\tt u}}
\newcommand{\Cf}{C_{\tt f}}
\newcommand{\Ns}{N_{\tt s}}
\newcommand{\Ts}{T_{\tt s}}
\newcommand{\tp}{\tau_{\tt p}}
\newcommand{\rp}{\rho_{\tt p}}
\newcommand{\rd}{\rho_{\tt d}}
\newcommand{\Rs}{R_{\tt s}}

\newcommand{\Rcinf}{\mathtt{R_{c, inf}}}
\newcommand{\Tr}{T_{\tt r}}
\newcommand{\sinrof}{\sinr_{o, {\tt fin}}}
\newcommand{\Na}{N_{\tt a}}
\newcommand{\lamr}{\lambda_{\tt r}}
\newcommand{\lamu}{\lambda_{\tt u}}

\def\nbg{{\mathbf{g}}}

\def\nbn{{\mathbf{n}}}
\def\nbo{{\mathbf{o}}}

\def\nbq{{\mathbf{q}}}
\def\nbr{{\mathbf{r}}}

\def\nbu{{\mathbf{u}}}

\def\nbw{{\mathbf{w}}}
\def\nbx{{\mathbf{x}}}
\def\nby{{\mathbf{y}}}

\def\nb0{{\mathbf{0}}}
\def\nb1{{\mathbf{1}}}

\def\nbI{{\mathbf{I}}}

\def\nbR{{\mathbf{R}}}

\def\nbW{{\mathbf{W}}}

\def\ncalB{{\mathcal{B}}}

\def\ncalP{{\mathcal{P}}}

\def\nbbC{{\mathbb{C}}}

\def\nbbE{{\mathbb{E}}}

\def\nbbP{{\mathbb{P}}}

\def\nrmh{{\rm h}}

\newtheorem{lemma}{Lemma}

\newtheorem{prop}{Proposition}

\newtheorem{remark}{Remark}

\def\R{\mathbb{R}}

\def\sinr{\mathtt{SINR}}			
\def\snr{\mathtt{SNR}}

\def\se{\mathtt{SE}}

\DeclareMathOperator{\Trc}{Tr}

\newcommand{\dP}[1]{\mathbb{P}\left[#1\right]}
\newcommand{\dE}[2]{\mathbb{E}_{#2}\left[#1\right]}

\newcommand{\cdf}{\mathtt{CDF}}
\newcommand{\pdf}{\mathtt{PDF}}

\allowdisplaybreaks 

\begin{document}
\title{Cell-Free Massive MIMO with Finite Fronthaul Capacity: A Stochastic Geometry Perspective}
\author{
Priyabrata Parida and Harpreet S. Dhillon
\thanks{Authors are with Wireless@VT, Dept. of Electrical and Computer Engineering, Virginia Tech, Blacksburg, VA (Email: \{pparida, hdhillon\}@vt.edu). The support of the US NSF (Grant ECCS-1731711) is gratefully acknowledged. This paper was presented in part at the IEEE Vehicular Technology Conference (VTC-Fall), 2018~\cite{Parida2018vtc}. A preliminary version of this work appears as a chapter in P. Parida's Ph.D. Thesis~\cite{pparidaPhDThesis}.} 
}

\maketitle

\vspace{-2cm}

\begin{abstract}
In this work, we analyze the downlink performance of a cell-free massive multiple-input-multiple-output system with finite capacity fronthaul links between the centralized baseband unit and the access point (APs). 
Conditioned on the user and AP locations, we first derive an achievable rate for a randomly selected user in the network that captures the effect of finite fronthaul capacity as a compression error. 
From this expression, we establish that for the traditional cell-free architecture where each AP serves all the users in the network, the achievable rate becomes zero as the network size grows. Hence, to have a meaningful analysis, for the traditional architecture, we model the user and AP locations as two independent binomial point processes over a finite region and provide an accurate theoretical result to determine the user rate coverage. 
For a larger (possibly infinite) network, we consider a user-centric architecture where each user in the network is served by a specified number of nearest APs that limits the fronthaul load.
For this architecture, we model the AP and user locations as two independent Poisson point processes (PPPs). Since the rate expression is a function of the number of users served by an AP, we statistically characterize the load in terms of the number of users per AP. 
As the exact derivation of the probability mass function of the load is intractable, we first present the exact expressions for the first two moments of the load. Next, we approximate the load as a negative binomial random variable through the moment matching method. 
Using the load results along with appropriate distance distributions of a PPP, we present an accurate theoretical expression for the rate coverage of the typical user. 
From the analyses we conclude that for the traditional architecture the average system sum-rate is a quasi-concave function of the number of users. Further, for the user-centric architecture, there exists an optimal number of serving APs that maximizes the average user rate.
\end{abstract}


\section{Introduction}
Cooperative cellular networks where a set of multiple base stations (BSs)/ access points (APs) simultaneously serve a set of users have been the subject of much investigation throughout the last decade. 
The latest incarnation of such networks is the cell-free massive multiple-input-multiple-output (mMIMO) systems that harness the benefits of network densification by extending the concept of cellular mMIMO to a distributed implementation~\cite{Ngo2017, Nayebi2017}.
In cell-free mMIMO networks, the APs perform a limited set of signal processing operations such as precoding/filtering using the local channel state information (CSI) while most of the baseband processing operations are carried out at the centralized baseband units (BBUs). 
The communication between the APs and BBUs is done through finite capacity fronthaul links.
One of the direct consequences of having finite fronthaul links is that the compression/quantization error gets introduced into the system, which affects the user performance. 
Hence, analyzing the network-wide performance of cell-free mMIMO with finite fronthaul capacity is an important requirement for the successful integration of this technology to the fifth generation (5G) and beyond networks. 
In this work, our goal is to model and analyze such a system using tools from stochastic geometry and provide useful system design guidelines.

\subsection{Related works}
We first discuss key prior works that focus on devising compression algorithms while taking into account the limited fronthaul capacity for other variants of cooperative cellular networks such as coordinated multipoint (CoMP) and cloud radio access networks (C-RAN). 
In~\cite{Simeone2009, Zakh2011}, authors provide information-theoretic insights regarding the capacity of a backhaul-constrained distributed MIMO system.
In other notable works, authors use optimization-based frameworks to devise compression algorithms that efficiently utilize the fronthaul capacity constraints while maximizing a certain performance metric (e.g., sum-rate) (cf. \cite{Park2013}, \cite{Zhang2013}). A comprehensive overview of such works can be found in~\cite{Park2014b, quek2017cloud}. While the insights obtained from these works are useful, they are not all directly applicable to a cell-free mMIMO system owing to its unique aspects such as beamforming based on local imperfect CSI at the APs as well as the time division duplex (TDD) mode of operation. A consequence of these aspects is a completely different user signal-to-interference-plus noise $(\sinr)$ expression compared to the system-level analyses of CoMP and C-RAN. This motivated a separate set of system-level analyses~\cite{Burr2018, Bashar2018,Bashar2019a,Femenias2019,Bashar2019b,Masoumi2020, Femenias2020,Bashar2021} for cell-free mMIMO with finite fronthaul capacity as briefly outlined below.

In~\cite{Burr2018}, the authors analyze the uplink performance using Bussgang decomposition to capture the effect of quantization error introduced due to finite fronthaul capacity. 
In particular, this work focuses on the effect of the number of quantization bits on the uplink outage probability.
In~\cite{Bashar2018}, the authors extend the framework of \cite{Burr2018} and compare the uplink performance of the scheme where both the quantized version of the received signal and quantized channel estimates are available at the BBU to the scheme where the BBU has the quantized weighted signal from each AP. In addition, an uplink max-min power allocation algorithm and an AP-user assignment scheme to reduce the fronthaul load are also proposed. 
In~\cite{Bashar2019b}, the authors compare the uplink performances of the following three cases:  perfect fronthaul links, when the quantized version of the estimated channel and signal available at the BBU, and when only quantized weighted signal is available at the BBU.
The uplink energy efficiency of cell-free network with finite fronthaul capacity is analyzed in~\cite{Bashar2019a}. 
In~\cite{Masoumi2020}, the authors study the performance of a cell-free network with hardware impairments where the authors compare the performance of three transmission strategies between the BBU and the APs through finite capacity links. 
The uplink and downlink performance of fronthaul constrained cell-free network with low resolutions ADCs is studied in~\cite{Femenias2020}.
Note that most of these works focus on {\em traditional} cell-free architecture where all the APs serve each user in the network. 
Since the user performance degrades with quantization/compression error, which depends on the number of users (load) per AP, each AP should ideally serve only a subset of users in the network. A network-centric approach that achieves this goal is proposed in~\cite{Bashar2018, Bashar2019b}. 
However, from the perspective of scalability and distributed implementation, a {\em user-centric} architecture is preferred where a user selects its set of serving APs~\cite{Buzzi2017a, Ngo2018, inter2019, Bursalioglu2019,bjornson2020, Buzzi2020, Bashar2021, demir2021}.
{To the best of our knowledge, the downlink performance of the user-centric cell-free architecture with finite fronthaul capacity has not been studied in the literature yet. Given its importance, one of our objectives is to bridge this gap in the literature.}

From the perspective of system-level analysis, a complementary approach to simulations-based studies is theoretical analyses using tools from stochastic geometry. 
To this end, there has been a lot of work that analyzes the performance of cooperative cellular networks, such as CoMP and C-RAN (cf.~\cite{baccelli2014st, Nigam2014, tanbourgi2014tr, lee2014sp, Govindasamy2018,Bao2018}). However, as mentioned earlier, the system architecture and key practical constrains of cell-free mMIMO, such as imperfect CSI, local beamforming, finite fronthaul capacity, result in a different $\sinr$ expression compared to the aforementioned works. Hence, the analyses developed in these works cannot be directly extended to the performance analysis of cell-free mMIMO system. 
Finally, while the performance of traditional cell-free architecture has been analyzed using stochastic geometry~\cite{Chen2017, Bursalioglu2019, Papaza2020,Kusaladharma2020}, none of them account for the finite fronthaul capacity, which is a key practical consideration and would yield a significantly different analysis. 
With this background on prior works, our contributions are outlined next.

\subsection{Contributions}
\subsubsection{System modeling} In this work, we consider the downlink of a cell-free mMIMO system with finite capacity fronthaul links. To capture the effect of finite fronthaul, we consider a point-to-point compression scheme between an AP and the BBU. Further, we focus on both the traditional cell-free mMIMO architecture, where each AP serves each user in the network, and a variant of the user-centric cell-free architecture. Since the compression error is a function of the number of users, the traditional cell-free network has to be of a finite size in order to limit the effect of compression error. Hence, for this architecture, we model the AP and user locations as two independent  binomial point processes (BPPs) that is in line with most of the works in the cell-free literature where fixed numbers of APs and users are considered. On the other hand, for the user-centric architecture, we are not restricted to consider a small network size. Therefore, we model the AP and user locations as two independent homogeneous Poisson point processes (PPPs) on $\R^2$ and assume that each user is served by a specified number of its nearest APs. We restrict our attention to conjugate beamforming. Conditioned on the AP and user locations, we derive an achievable rate expression for a randomly selected user that captures the effect of finite fronthaul capacity in both the architectures.

\subsubsection{Load characterization of user-centric architecture}
Due to the dependence of compression error on the number of users served by an AP, the statistics of the load in terms of the number of users is important for the system-level analysis.  While this number of fixed for the traditional architecture, the load is a function of user and AP densities as well as the number of APs that serve a given user in the user-centric architecture. 
Hence, for the latter, we first determine the load distribution for the set of tagged APs that serve the typical user. Since an exact determination of the probability mass function $(\pmf)$ of the number of users associated with each tagged AP is intractable, we derive the first two moments of the load and then approximate load for each of the tagged APs as a negative binomial random variable through the moment matching method. This result is later used to derive the rate coverage of the typical user in the user-centric architecture. Further, we use a similar methodology to derive the load result for the typical AP in the network. This result is useful in network dimensioning, especially determining the desired capacity of the fronthaul link between the typical AP and the BBU to satisfy a certain signal to compression noise ratio $({\tt SCNR})$. It is worth mentioning that this result has a direct equivalence to the degree distribution in an {\em AB random geometric graph}.

\subsubsection{Performance analysis of the traditional architecture}
Using the above results, we first derive the downlink user rate coverage result for the traditional architecture. Leveraging the relevant distance distributions for a BPP, we provide an approximate expression to analytically evaluate the rate coverage averaged over the AP and user locations.
From our analyses, we infer that the average system sum-rate is a strictly quasi-concave function of the number of users, and the optimal number of users to achieve the maximum system sum-rate increases with increasing fronthaul capacity.
Further, in contrast to the established notion that fully distributed MIMO is superior to the collocated MIMO, our results suggest that in the presence of high-quality CSI at the APs, a less distributed form of cell-free mMIMO is better, i.e., for an equal number of antennas in the system, it is better to deploy a fewer APs with more antennas per AP.

\subsubsection{Performance analysis of the user-centric architecture}
Using the load distribution result of the typical AP, we highlight the interplay between key system parameters such as the fronthaul capacity, the ${\tt SCNR}$, and the number of serving APs. 
Further, exploiting the statistical properties associated with a PPP along with a few subtle approximations, we derive the rate coverage result for the typical user. 
A key ingredient of this derivation is the load distribution results for the tagged APs.
All the theoretical results are validated through extensive Monte Carlo simulations.

\section{System Model}\label{sec:SysMod}

\begin{figure*}[!htb]
\centering
\begin{subfigure}{0.45\textwidth}
  \centering
  \includegraphics[width=\linewidth]{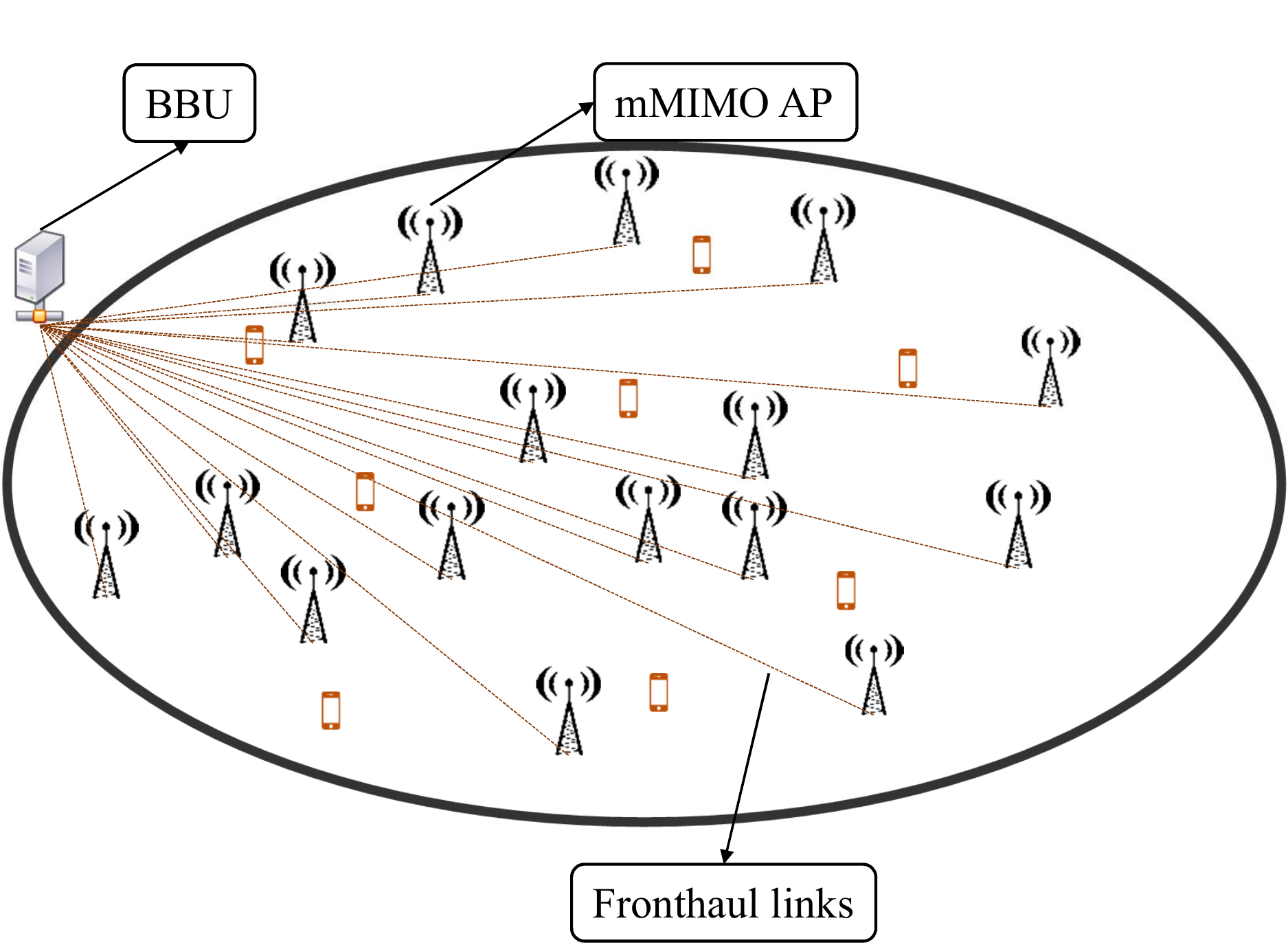}
\end{subfigure}
\begin{subfigure}{0.45\textwidth}
  \centering
  \includegraphics[width=\linewidth]{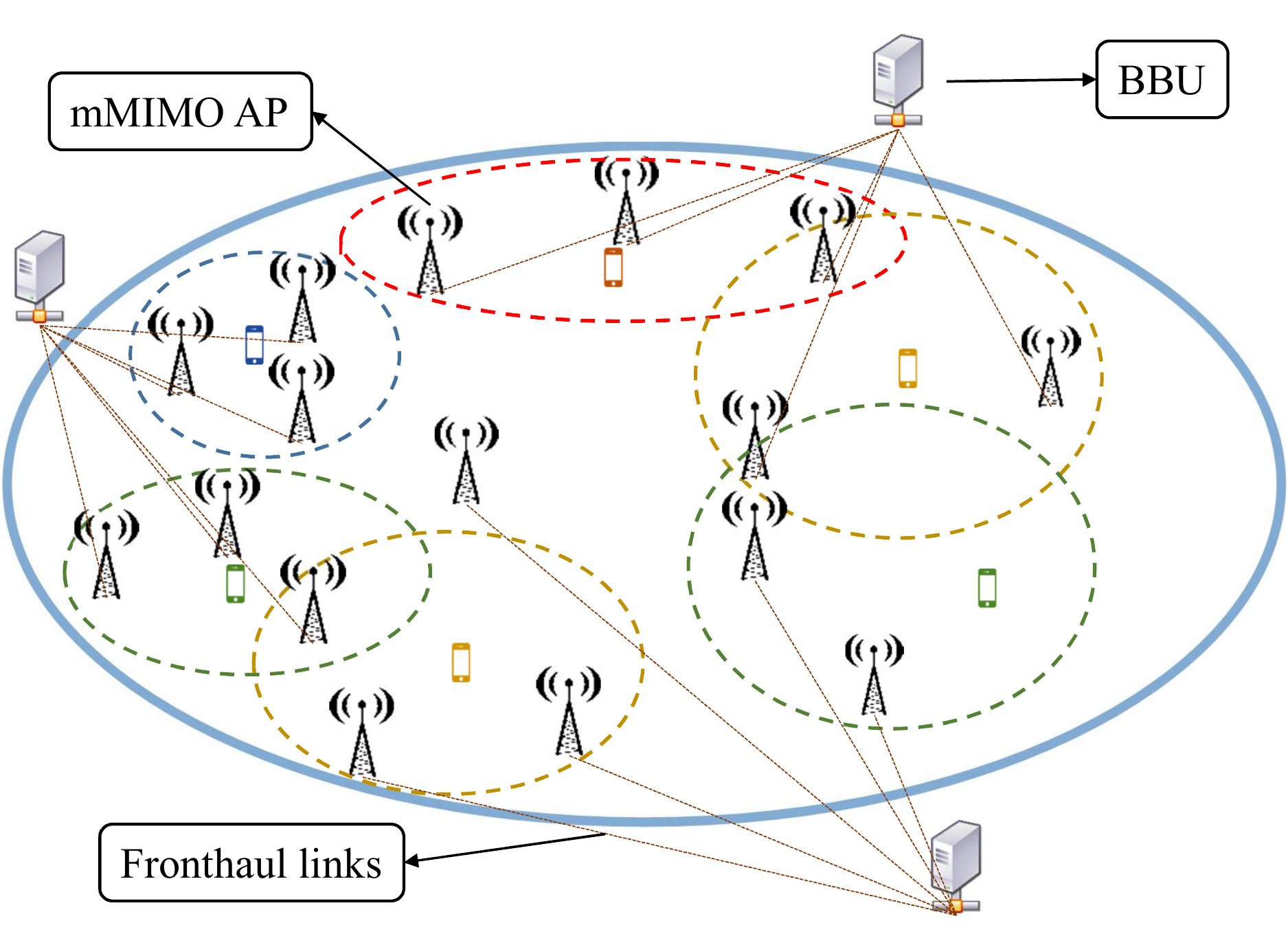}
\end{subfigure}%
\caption{\small Representative network diagrams: (left) the traditional architecture, where each AP serves all the user in the network, and (right) the user-centric architecture, where each user is served by  its nearest three APs as marked by dotted circles.}
\label{fig:DiMIMO}
\end{figure*}

We limit our attention to the downlink of a cell-free mMIMO system.
The sets of AP and user locations are given by $\fir$ and $\fiu$, respectively. 
To capture the spatial randomness in the AP and user locations, we model $\fir$ and $\fiu$ by appropriate point processes. The corresponding discussions on the point processes are relegated to the following sections as it is not necessary for the results derived in this section. 
We assume that each AP has $\Na$ antennas.
The distance between a user at $\nbu_k \in \fiu$ and an AP at $\nbr_m \in \fir$ is denoted by $d_{mk}$.
All the APs are connected to a BBU through a fronthaul network, where the capacity of each link is $\Cf$ bits/s/Hz.
As mentioned earlier, in case of the traditional cell-free architecture, all the APs serve all the users in the network.
In contrast, in case of the user-centric network architecture, we consider that each user is served by its nearest $\Ns$ APs.
Both the architectures are illustrated in Fig.~\ref{fig:DiMIMO}.

\subsection{Compression at the BBU}
Due to limited fronthaul capacity, the BBU employs a lossy compression scheme to forward user symbols to the APs.
Let an AP at $\nbr_o$ serves a set of $K_o$ users $\Phi_{uo} \subseteq \fiu$. Note that in the case of traditional architecture, $\Phi_{uo} = \fiu$.
Let $\nbq_o = [q_{1_o}, q_{2_o}, \ldots, q_{K_o}]^T$ be the signal vector consisting of the symbols to be transmitted to the users in $\Phi_{uo}$.
We consider that $\nbq_o$ is a circularly symmetric complex Gaussian random vector and $\nbq_o \sim {\cal CN}(\boldsymbol{0}_{K_o}, \rho_{q_o} \nbI_{K_o})$, where $\rho_{q_o} = \dE{|q_{1_o}|^2}{} = \dE{|q_{2_o}|^2}{} = \ldots = \dE{|q_{K_o}|^2}{}$, $\boldsymbol{0}_{K_o}$ denotes a $K_o \times 1$ all zero vector, and $\nbI_{K_o}$ denotes a $K_o \times K_o$ identity matrix.
Using a lossy compression scheme, the BBU transmits $\hat{\nbq}_o = [\hat{q}_{1_o}, \hat{q}_{2_o}, \ldots, \hat{q}_{K_o}]^T$  over the fronthaul links to the AP.
Similar to \cite{Simeone2009}, we consider $\hat{\nbq}_o = \nbq_o + \tilde{\nbq}_o$, where $\tilde{\nbq}_o \sim {\cal CN}(\boldsymbol{0}_{K_o}, \rho_{\tilde{q}_o} \nbI_{K_o})$ is the compression error vector and $\rho_{\tilde{q}_o} = \dE{|\tilde{q}_{1_o}|^2}{} = \dE{|\tilde{q}_{2_o}|^2}{} = \ldots = \dE{|\tilde{q}_{K_o}|^2}{}$.
Further, we assume that $\nbq_o$ and $\tilde{\nbq}_o$ are uncorrelated. 
Since both are Gaussian random vectors, they are independent as well.
From the above exposition, it is clear that $\hat{\nbq} \sim {\cal CN}(\boldsymbol{0}_{K_o}, (\rho_{\tilde{q}_o} + \rho_{q_o}) \nbI_{K_o})$.
If $\dE{|\hat{q}_{k_o}|^2}{}$ is the same for all $k = 1, 2, \ldots K_o$, then both $\rho_{\tilde{q}_o},  \rho_{q_o}$ depend on the fronthaul capacity $\Cf$, as discussed in the following lemma.

\begin{lemma}
For a fronthaul capacity $\Cf$ and number of users $K_o$ served by the typical AP, $\rho_{q_o} = \left(1- 2^{-\Cf/K_o}\right) \dE{|\hat{q}_{k_o}|^2}{}$ and $\rho_{\tilde{q}_o} = 2^{-\Cf/K_o} \dE{|\hat{q}_{k_o}|^2}{}$.
\end{lemma} 
\begin{IEEEproof}
The amount of information that can be transmitted from the BBU to each AP is upper bounded by the fronthaul capacity $\Cf$.
Hence, we write 
\begin{align*}
I(\hat{\nbq}_o; \nbq_o) \leq \Cf \implies & \nrmh(\hat{\nbq}_o) - \nrmh(\hat{\nbq}_o|\nbq_o) \leq \Cf
\implies \sum_{i=1}^{K_o} \nrmh(\hat{q}_{i_o}) - \sum_{i=1}^{K_o}\nrmh(\hat{q}_{i_o}|{q}_{i_o}) \leq \Cf \\
\implies & \log_2(\pi e (\rho_{q_o} + \rho_{\tilde{q}_o})) -  \log_2(\pi e \rho_{\tilde{q}_o}) \leq \frac{\Cf}{K_o},
\end{align*}
where $I(x;y)$ denotes the mutual information between two random variables $x$ and $y$, $\nrmh(x)$ denotes the differential entropy of a random variable $x$, and the last step follows from the fact that $\hat{q}_i$s and $\tilde{q}_i$s are complex Gaussian random variables.
Ideally, the BBU would like to transmit the maximum information. Hence, we get
\begin{align*}
\log_2\left(1 + \frac{\rho_{q_o}}{\rho_{\tilde{q}_o}}\right) = \frac{\Cf}{K_o} \implies  \frac{\rho_{q_o}}{\rho_{\tilde{q}_o}} = 2^{\Cf/K_o}-1.
\end{align*}
The expression in the lemma follows directly using the fact that $\rho_{q_o} + \rho_{\tilde{q}_o} = \dE{|\hat{q}_{k_o}|^2}{}$.
If we consider that $\dE{|\hat{q}_{k_o}|^2}{} = 1$, then $\rho_{q_o} = \left(1- 2^{-\Cf/K_o}\right)$ and $\rho_{\tilde{q}_o} = 2^{-\Cf/K_o}$.
\end{IEEEproof}

\begin{remark}\label{rem:MaxLoad}
The ${\tt SCNR}$, defined as $ \frac{\rho_{q_o}}{\rho_{\tilde{q}_o}} = 2^{\Cf/K_o}-1$, is a decreasing function of the number of users served by the AP. While in the case of the traditional cell-free mMIMO, the ${\tt SCNR}$ can only be improved by increasing $\Cf$, in the case of user-centric architecture, ${\tt SCNR}$ can also be improved by limiting the maximum number of users  that should be scheduled by the typical AP. Hence, for a given $\Cf$ and target ${\tt SCNR}$ threshold $\Ts$, the maximum number of scheduled users $K_{\tt max}$ should satisfy $K_{\tt max} \log_2(1 + \Ts) \leq \Cf$.
\end{remark}

\subsection{Uplink channel estimation}\label{subsec:ULChanEst}
Let $\mathbf{g}_{mk} = \sqrt{\beta_{mk}} \mathbf{h}_{mk}$ be the channel gain between the AP at $\nbr_m$ and the user at $\nbu_k$, where $\beta_{mk}$ captures the large-scale channel gain and $\mathbf{h}_{mk} \sim {\cal CN}(\mathbf{0}_{\Na}, \mathbf{I}_{\Na})$ captures the small-scale channel fluctuations. We consider that $\beta_{mk}$ is only due to the distance dependent path loss, i.e., $\beta_{mk} =  l(d_{mk})^{-1}$, where $l(\cdot)$ is a non-decreasing  path loss function presented in Section~\ref{sec:Results}. 

In order to obtain the channel estimates, we consider that each user uses a pilot from a set of $P$ orthogonal pilot sequences of $\tp$ symbol duration, which is assumed to be less than the coherence interval.
Further, the transmit signal-to-noise ratio ($\snr$) of each symbol in a pilot is $\rp$.
Since we assume that these $P$ sequences are orthogonal to each other, $\tp \geq P$ and $\boldsymbol{\psi}_i^H\boldsymbol{\psi}_j = \mathbf{1}(i = j)$, where $\mathbf{1}(\cdot)$ denotes the indicator function.
Let the pilot used by the user at $\nbu_k$ be $\boldsymbol{\psi}(k)$.
During the pilot transmission phase, the received signal matrix $\nbR_o \in \nbbC^{\Na\times \tp}$ at the typical AP is
\begin{align*}
\nbR_o = \sqrt{\tp \rp} \sum_{\nbu_k \in \fiu} \nbg_{ok} \boldsymbol{\psi}(k)^T + \nbW_o ,
\end{align*}
where each element of $\nbW_o$ is ${\cal CN}(0, 1)$.
Let $\hat{\nbg}_{ok}$ be the estimated channel vector at the AP $\nbr_o$ for the user $\nbu_k \in \fiu$ that is obtained after performing minimum-mean-squared-error (MMSE) channel estimation. 
Further, $\tilde{\nbg}_{ok}$ be the estimation error vector. Using the properties of MMSE estimation~\cite{Nayebi2017}, we write $
\hat{\nbg}_{ok} \sim {\cal CN}\left(\mathbf{0}_{\Na}, \gamma_{ok}\mathbf{I}_{\Na}\right)$ and $\tilde{\nbg}_{ok} \sim {\cal CN}\left(\mathbf{0}_{\Na}, \left(\beta_{ok} - \gamma_{ok}\right)\mathbf{I}_{\Na}\right),
$
where 
\begin{align*} 
\gamma_{ok} = \frac{\tp \rp \beta_{ok}^2}{1 + \tp \rp \sum_{\nbu_j \in \Phi_{u}} \boldsymbol{\psi}(k)^H \boldsymbol{\psi}(j) \beta_{oj}}. \numberthis
\label{eq:EstVariance}
\end{align*}

\subsection{Downlink data transmission}\label{Subsec:DLTx}
In this work, we consider that each AP employs conjugate beamforming based on the local CSI.
Hence, the precoded symbol transmitted by the AP at $\nbr_o$ is given as 
\begin{align*}
\nbx_o = \sum_{\nbu_i \in \Psi_{uo}} \sqrt{\rd \eta_{oi}} \frac{\hat{\nbg}_{oi}^*}{\sqrt{\nbbE[\|\hat{\nbg}_{oi}\|^2]}} \hat{q}_{i_o} =  \sum_{\nbu_i \in \Psi_{uo}} \sqrt{\rd \eta_{oi}} \nbw_{oi} \hat{q}_{i_o},
\end{align*}
where $\rd$ is the downlink transmit $\snr$, $\eta_{oi}$ is normalization coefficient used by the typical AP for the user at $\nbu_i$ to satisfy the average power constraint $
\Trc(\dE{\nbx_o \nbx_o^H}{}) \leq \rd$,
and $\Psi_{uo} \subseteq \Phi_{uo}$ is the set of scheduled users associated with the AP at $\nbr_o$ such that $|\Psi_{uo}| \leq K_{\tt max}$ for the user-centric architecture.
Note that for the traditional architecture, $\Psi_{uo}  = \Phi_{uo} = \fiu$ and $K_{\tt max} = K_o$. 
We observe that by setting $\eta_{mk} = 1/K_{\tt max}$ and $\rho_{\hat{q}_o} = \nbbE[|\hat{q}_{i_o}|^2] = 1$, the above constraint is satisfied. More sophisticated power allocation algorithms, such as max-min power allocation, can of course be considered. 
However, the equal power allocation scheme has its own advantages of lower complexity and admitting a distributed implementation.  Besides, this scheme provides a degree of tractability in the coverage analysis as will be evident in the sequel. 

\subsection{An achievable rate for a randomly selected user}
Now, we present an achievable rate for a randomly selected user in the network that is applicable for both types of architectures.
Consider that a randomly selected user is located at $\nbu_o$ and is served by the set of APs $\Phi_{ro} \subseteq \fir$.
The received signal at this user is given as 
\begin{align*}
y_o^{\tt dl} = & \sum_{\nbr_l \in \Phi_{ro}} \nbg_{lo}^T \nbx_l + \sum_{\nbr_j \in \Phi_{ro}^C} \nbg_{jo}^T \nbx_j + \nbn_o  \\
 = & \sum_{\nbr_l \in \Phi_{ro}} \! \! \sqrt{\rd \eta_{lo}} \frac{\nbg_{lo}^T \hat{\nbg}_{lo}^*}{\sqrt{\Na\gamma_{lo}}} \hat{q}_{lo}  +  \! \!\sum_{\nbr_l \in \Phi_{ro}} \! \sum_{\nbu_i \in \tilde{\Psi}_{ul}} \!\! \frac{\sqrt{\rd \eta_{li}} \nbg_{lo}^T \hat{\nbg}_{li}^*}{\sqrt{\Na\gamma_{li}}} \hat{q}_{li} + \!\! \sum_{\nbr_l \in \Phi_{ro}^C} \! \sum_{\nbu_i \in {\Phi}_{ul}} \!\! \frac{\sqrt{\rd \eta_{li}} \nbg_{lo}^T \hat{\nbg}_{li}^*}{\sqrt{\Na\gamma_{li}}} \hat{q}_{li} + n_o,
\numberthis
\label{eq:ro}
\end{align*}
where $\Phi_{ro}^C = \fir \setminus \Phi_{ro}$ and $\tilde{\Psi}_{ul} = {\Psi}_{ul} \setminus \nbu_o$.
In the following lemma, we provide an expression for an achievable rate (a lower bound on capacity). Note that in favor of a simpler exposition, we ignore the constant pre-log factors such as bandwidth and fraction of downlink transmission duration in a TDD setup as we do not study the corresponding trade offs in this work.
\begin{lemma}\label{lem:RoLB}
Conditioned on $\fir$ and $\fiu$, an achievable rate of the typical user at $\nbu_o$ is given as
\begin{align*}
\se_{o} = \log_2\left(1 + \sinr_o\right) \quad \text{bits/s/Hz}, \quad \text {where}
 \numberthis
\label{eq:RoLB}
\end{align*}
\begin{align}
\sinr_o =  \frac{\rd \Na\left(\sum\limits_{\nbr_l \in \Phi_{ro}} \sqrt{ \frac{\gamma_{lo} (1 - 2^{-\Cf/k_l})}{K_{\tt max}}}\right)^2}{\rd \Na\! \sum\limits_{\nbr_l \in \Phi_{ro}} \! {\gamma_{lo}} \frac{2^{-\Cf/k_l}}{K_{\tt max}} + \rd \sum\limits_{\nbr_l \in \Phi_{r}} \beta_{lo} + \rd \Na\sum\limits_{\nbu_i \in \{\ncalP_o \setminus \nbu_o\}} \left(\sum\limits_{\nbr_l \in \Phi_{ri}} \sqrt{\frac{\gamma_{lo}}{K_{\tt max}}}\right)^2 + 1}
\label{eq:SINR_LB_Gen}
\end{align}
and $\ncalP_o$ is the set of users that use the same pilot sequence as the typical user $\nbu_o$, and $k_l$ is the number of users served by the AP at $\nbr_l$.
\end{lemma}
\begin{IEEEproof}
Please refer to Appendix~\ref{app:LB}.
\end{IEEEproof}

\section{Rate Coverage for traditional cell-free mMIMO}\label{sec:RateTrad}
In this section, we derive the rate coverage result for the traditional cell-free mMIMO system where each AP serves all the users in the network.
If we consider an infinite network on $\R^2$, then as per the result of Lemma~1, the ${\tt SCNR} \rightarrow 0$ as $K_o \rightarrow \infty$ and subsequently $\sinr_o\rightarrow 0$ as given in Lemma~\ref{lem:RoLB}.
Hence, for a meaningful analysis of the traditional architecture, we need to consider a finite network, e.g., a shopping mall.
Therefore, we assume the system is limited to $\ncalB_{\Rs}(\nbo)$, a finite circular region of radius $\Rs$ centered at $\nbo$, where the set of APs $\fir = \{\nbr_1, \nbr_2, \ldots, \nbr_M\}$ are randomly and uniformly distributed.
Further, $\fiu = \{\nbu_1, \nbu_2, \ldots, \nbu_{K_o}\}$ contains the set of user locations that are uniformly and randomly distributed in $\ncalB_{\Rs}(\nbo)$ and are independent of AP locations. 
{Note that by construction, $\fir$ and $\fiu$ form two independent BPPs.
Alternatively, one can consider modeling $\fir$ and $\fiu$ as two independent PPPs over a finite region. However, in that case $K_o$ becomes a Poisson random variable that theoretically has an infinite support. Hence, in certain cases when $K_o$ is high, $\sinr_o$ becomes undesirably low for a fixed $\Cf$. Therefore, we consider a fixed number of users and APs in the network by modeling them as BPPs. This is also consistent with most of the studies in the cell-free mMIMO literature that also consider a fixed number of users and APs in the network.}
As assumed in the cell-free mMIMO literature, we consider that $M \Na\gg K_o$. 
Further, we assume that the coherence block is sufficiently long to ensure that $\tp \geq K_o$. As a consequence, pilots are not reused in the network, thereby eliminating the effect of pilot contamination. 
Under these assumptions, using the result of Lemma~\ref{lem:RoLB}, the achievable rate of a user at $\nbu_o$  is given as $\se_{o, {\tt fin}} = \log_2\left(1 + \sinr_{o, {\tt fin}}\right)$, where 
\begin{align}
\sinr_{o, {\tt fin}} = \frac{\rd \frac{\Na}{K_o} (1 - 2^{-\Cf/K_o})\left(\sum\limits_{m=1}^M  \sqrt{\gamma_{mo}}\right)^2}{\rd\frac{\Na}{K_o} 2^{-\Cf/K_o} \sum\limits_{m=1}^M \gamma_{mo} + \rd \sum\limits_{m=1}^M \beta_{mo} + 1}
\label{eq:SINR_LB_JS}
\end{align}
is the $\sinr$ of the user at $\nbu_o$ in this finite network.

Our goal is to determine the rate coverage $\mathtt{R_{c, fin}}(\Tr) = \nbbP[\log_2( 1 + \sinrof) > \Tr]$ for a randomly selected user that requires averaging over the distances of the APs from the user. 
Hence, now we present a few relevant distance distributions for a BPP. 
\subsection{Relevant distance distributions in a BPP}
Let $R_o$ be the distance of the user at $\nbu_o$ from the center of the circle ${\cal B}_{\Rs}(\nbo)$. 
Since $\nbu_o$ is uniformly and randomly distributed in ${\cal B}_{\Rs}(\nbo)$, the cumulative distribution function ($\cdf$) and probability density function ($\pdf$) of $R_o$ is given as 
\begin{align*}
& F_{R_o}(r) = {r^2}/{\Rs^2}, \quad \text{and} \quad f_{R_o}(r) = {2r}/{\Rs^2}, \quad 0 \leq r \leq \Rs.\numberthis
\label{eq:PDFRo}
\end{align*}
Next, we present the distance distribution between $\nbu_o$ to a randomly distributed AP in ${\cal B}_{\Rs}(\nbo)$.
\begin{lemma}\label{lem:Dmo}
Conditioned on the distance $R_o$, the $\cdf$ of the distance between the user at $\nbu_o$ and the AP at $\nbr_m$ is given as
\begin{align*}
F_{D_{mo}}(d|r_o) = \frac{d^2}{\Rs^2} \ \mathbf{1}_{0 \leq d < \Rs- r_o} +  \bigg(\frac{d^2\left(\theta^* - \frac{\sin(2 \theta^*)}{2}\right)}{\pi \Rs^2}  + \frac{\left(\phi^*-\frac{\sin(2 \phi^*)}{2}\right)}{\pi} \bigg)\mathbf{1}_{\Rs - r_o \leq d \leq \Rs + r_o},
\end{align*}
and corresponding $\pdf$ is given as 
\begin{align*}
f_{D_{mo}}(d|r_o) = & \frac{2d}{\Rs^2} \ \mathbf{1}_{0 \leq d < \Rs- r_o}  +  \frac{2d\theta^*}{\pi \Rs^2} \ \mathbf{1}_{\Rs - r_o \leq d \leq \Rs + r_o}
\end{align*}
where $\theta^* = \arccos\left(\frac{d^2 + r_o^2 - \Rs^2}{2 r_o d}\right), \phi^* = \arccos\left(\frac{\Rs^2 + r_o^2 - d^2}{2 r_o \Rs}\right)$, and $\mathbf{1}_{(\cdot)}$ is the indicator function.
\end{lemma}
\begin{IEEEproof}
We provide a proof sketch for this lemma. Please refer to \cite[Lemma~1]{AfsDhi2017} for the detailed proof. Without loss of generality, consider that $\nbu_o = (r_o, 0)$. Then, conditioned on $\nbu_o$ (equivalently $r_o$), a uniformly distributed point in ${\cal B}_{\Rs}(\nbo)$ can lie either in the circle ${\cal B}_{\Rs-r_o}(\nbu_o)$ or in the region ${\cal B}_{\Rs}(\nbo) \setminus {\cal B}_{\Rs-r_o}(\nbu_o)$. In the $\cdf$ expression of the lemma, both these conditions are  captured by the indicator function and corresponding conditional $\cdf$s are presented. The expression for the $\pdf$ is obtained by taking the derivative of the $\cdf$ with respect to $d$ along with some algebraic manipulation.
\end{IEEEproof}
Now, using the results from order statistics, we present the conditional distance distribution between $\nbu_o$ and its nearest AP.
\begin{lemma}\label{lem:Doo}
Conditioned on the distance $R_o$, the $\cdf$ of the distance $D_{oo}$ between $\nbu_o$ and its nearest AP is given as $F_{D_{oo}}(d_{oo}|r_o) = \dP{D_{oo} \leq d_{oo}|r_o}{} =  1 - (1 - F_{D_{mo}}(d_{oo}|r_o))^{M},
$
and the corresponding $\pdf$ is given as 
$
f_{D_{oo}}(d_{oo}|r_o) =   M f_{D_{mo}}(d_{oo}|r_o) (1 - F_{D_{mo}}(d_{oo}|r_o))^{M-1},
$
where $f_{D_{mo}}, F_{D_{mo}}$ were presented in Lemma~\ref{lem:Dmo}.
\end{lemma}
Note that conditioned on the distance $D_{oo}$, rest of the APs in ${\cal B}_{\Rs}(\nbo)$ are uniformly and randomly located in ${\cal B}_{\Rs}(\nbo) \setminus {\cal B}_{d_{oo}}(\nbu_o)$, where $d_{oo}$ is a realization of $D_{oo}$.
In the following lemma, we present the distribution of the distance between a randomly located AP in the above region and $\nbu_o$.
\begin{lemma}\label{lem:hatDmo}
Conditioned $D_{oo}$ and $R_o$, the $\pdf$ of the distance $\hat{D}_{mo}$ between a randomly located AP in ${\cal B}_{\Rs}(\nbo) \setminus {\cal B}_{d_{oo}}(\nbu_o)$ and $\nbu_o$ is given as 
\begin{align*}
f_{\hat{D}_{mo}}(d|d_{oo}, r_o) = \frac{f_{D_{mo}}(d|r_o)}{1 - F_{D_{mo}}(d_{oo}|r_o)}, \quad d_{oo} \leq d \leq r_o + \Rs.
\end{align*}
\end{lemma}
\begin{IEEEproof}
We provide a proof sketch for this lemma. For the detailed proof, please refer to \cite[Lemma~3]{AfsDhi2017}. Conditioned on $D_{oo}$, rest of the APs are uniformly distributed in ${\cal B}_{\Rs}(\nbo) \setminus {\cal B}_{d_{oo}}(\nbu_o)$. Hence, the distribution of the distance $\hat{D}_{mo}$ follows the lower truncated distribution of $D_{mo}$, which is captured in the above expression.
\end{IEEEproof}
Next, using the above distance distributions, we derive the rate coverage expression.

\subsection{Approximate evaluation of average achievable user rate}
The exact evaluation of rate coverage $\mathtt{R_{c, fin}}$ is challenging as it requires an $(M+1)$-fold integration to average it over the locations of all the $M$ APs and the user at $\nbu_o$.
Notice that the $\sinrof$ in  \eqref{eq:SINR_LB_JS} has the following terms:
\begin{align*}
I_1 = \sum_{m=1}^M \sqrt{\gamma_{mo}},\quad I_2 = \sum_{m=1}^M \gamma_{mo}, \quad I_3 = \sum_{m=1}^M \beta_{mo}. \numberthis
\label{eq:DefTerms}
\end{align*}
Since there is no pilot contamination, $\gamma_{mk}(d_{mk}) = \frac{\tp \rp l(d_{mk})^{-2}}{1 + \tp \rp l(d_{mk})^{-1}}$. Further, $\gamma_{mk}$ is a decreasing function of $d_{mk}$.
{Due to path loss, these terms are dominated by contributions from a few nearest APs.
Hence, we approximate $I_1, I_2,$ and $I_3$ as the sum of exact contribution from the {nearest} AP and the mean contribution from the rest of the APs conditioned on the distance $d_{oo}$ between $\nbu_o$ and its nearest AP. Hence, we write
\begin{align*}
I_1(d_{oo}, r_o) & \approx \sqrt{\gamma_{oo}} + \nbbE\bigg[\sum\limits_{{m=1, m\neq o}}^M \sqrt{\gamma_{mo}}\bigg| d_{oo}, r_o\bigg], \quad I_2(d_{oo}, r_o) \approx {\gamma_{oo}} + \nbbE\bigg[\sum\limits_{{m=1, m\neq o}}^M {\gamma_{mo}}\bigg| d_{oo}, r_o\bigg], \\
I_3(d_{oo}, r_o) & \approx {\beta_{oo}} + \nbbE\bigg[\sum\limits_{{m=1, m\neq o}}^M {\beta_{mo}}\bigg| d_{oo}, r_o\bigg].\numberthis
\label{eq:DomApprox}
\end{align*}
It is worth mentioning that this approach based on dominant interferers has been extensively used in the stochastic geometry literature (cf.~\cite{Madhu2014, chetlur2017}).}
Note that conditioned on $D_{oo}$, distances between $\nbu_o$ and rest of the APs in the network are i.i.d. Hence, using Campbell's theorem, \eqref{eq:DomApprox} can be written as 
\begin{align*}
\hat{I}_1(d_{oo}, r_o) & = \sqrt{\gamma_{oo}} + (M-1) \int_{r= d_{oo}}^{r_o + \Rs} \frac{\sqrt{\tp \rp} l(r)^{-1}}{\sqrt{1 + \tp \rp l(r)^{-1}}} f_{\hat{D}_{mo}}(r|d_{oo}, r_o) {\rm d}r, \\
\hat{I}_2(d_{oo}, r_o) & = \gamma_{oo} + (M-1) \int_{r= d_{oo}}^{r_o + \Rs} \frac{{\tp \rp} l(r)^{-1}}{{1 + \tp \rp l(r)^{-1}}} f_{\hat{D}_{mo}}(r|d_{oo}, r_o) {\rm d}r, \\
\hat{I}_3(d_{oo}, r_o) & = \beta_{oo} + (M-1) \int_{r= d_{oo}}^{r_o + \Rs} l(r)^{-1} f_{\hat{D}_{mo}}(r|d_{oo}, r_o) {\rm d}r. \numberthis
\label{eq:SinrAprxTerms}
\end{align*}

With the above approximation, in the next Proposition, we present an expression to evaluate the rate coverage of the typical user in this finite cell-free mMIMO network.
\begin{prop}\label{prop:RcFinite}
For a given threshold $\Tr$, the rate coverage of a randomly selected user in the network is given as 
\begin{align*}
\mathtt{R_{c, fin}}(\Tr) & = \int_{r_o = 0}^{\Rs}  \int_{d_{oo}=0}^{\Rs} \mathbf{1}\left(\sinr_{o}^{\mathrm{Apx}}(d_{oo}, r_o) > 2^{\Tr}-1\right) f_{D_{oo}}(d_{oo} | r_o) f_{R_o}(r_o) {\rm d}d_{oo} {\rm d}r_o,
\end{align*}
where 
\begin{align}
\sinr_{o}^{\mathrm{Apx}}(d_{oo}, r_o) =
 \frac{\rd \frac{N}{K_o} (1 - 2^{-\Cf/K_o}) (\hat{I}_1(d_{oo}, r_o))^2}{\rd \frac{N}{K_o} (\hat{I}_2(d_{oo}, r_o))  2^{-\Cf/K_o} + \rd \hat{I}_3(d_{oo}, r_o) + 1},
\label{eq:RoLBApx}
\end{align}
and the $\pdf$s of $D_{oo}$ and $R_o$ are presented in Lemma~\ref{lem:Doo} and \eqref{eq:PDFRo}, respectively.
\end{prop}
\begin{IEEEproof}
This result follows by first replacing different terms in the $\sinrof$ by their approximations given in \eqref{eq:SinrAprxTerms} to obtain $\sinr_{o}^{\mathrm{Apx}}(d_{oo}, r_o)$. In the next step, we decondition over $D_{oo}$ and $R_o$ to obtain $\mathtt{R_{c, fin}}(\Tr)$.
\end{IEEEproof}

The result of the above proposition concludes the rate coverage derivation for a traditional cell-free mMIMO system with finite fronthaul capacity. 
Next, we focus on the user-centric cell-free mMIMO.
In this case, since we are not restricted to a finite network size, we model $\fir$ and $\fiu$ as two independent homogeneous PPPs on $\mathbb{R}^2$.
The densities of $\fir$ and $\fiu$ are $\lamr$ and $\lamu$, respectively.
Further, each user is served by its nearest $\Ns$ APs. Observe that the $\sinr$ expression of Lemma~2 is a function of the number of users served by each AP in the network. Owing to the spatial randomness of both user and AP locations, the number of users served by each AP is a random variable. Therefore, to derive the rate coverage expression, we need the statistical properties of the load associated with an AP  that are presented in the next section.

\section{Load Characterization in User-Centric Architecture}\label{sec:LoadDerivation}
Before proceeding further, we need to provide the distinction between the {\em typical AP} and the set of {\em tagged APs} in the network. 
The typical AP is by definition a randomly selected AP in $\fir$. On the other hand the set of tagged APs are the serving APs of the typical user, which is selected randomly from $\fiu$.
This random selection of the typical user makes it more likely to be served by APs that have larger service regions. 
This effect is reminiscent of the {\em waiting time} paradox in queuing systems and the difference between $0$-cell and the typical cell in a Poisson-Voronoi tessellation~\cite{mecke1999re, Mankar2020}. 
Since we are focusing on the performance analysis of the typical user, we need the load distribution result associated with the set of tagged APs. 
On the other hand, from network dimensioning perspective, such as provisioning of fronthaul capacity, we also need the statistical information on the load associated with the typical AP. 
In the next subsection, we first derive the load distribution results associated with the set of tagged APs.

\subsection{The load of a tagged AP}\label{subsec:TagAP}
The statistical metric that we are interested in is the $\pmf$ of the load. The exact derivation of $\pmf$ is intractable. Hence, we first derive the exact result for the first two moments of the number of users for a tagged AP. Then we approximate the load though an appropriate random variable using the moment matching method. 

\subsubsection{Determination of the first two moments}
Since $\fiu$ is a homogeneous PPP, it is translation invariant. Hence, we assume that the typical user is located at the origin $\nbo$. 
It is worth mentioning that the loads associated with each of the tagged APs are not identical. 
Hence, we need to present a generic result that is a function of the serving AP rank in terms of the distance from the typical user.
Next, we derive the first two moments of the load for the $N$-th nearest tagged AP. 
\begin{lemma}\label{lem:LoadTagRRH}
The first moment of the number of users (excluding the typical user) served by the $N$-th nearest AP to the typical user at the origin is given as
\begin{align*}
\nbbE[K_{N}] = 2 \pi \lamr \lamu \int_{r_o= 0}^{\infty} {\rm d}r_o \int_{d_x=0}^{\infty} {\rm d}d_x \int_{v_x=0}^{2 \pi}{\rm d}v_x \ h_{\tt tag, m_1}(r_o, d_x, v_x, N)   d_x   r_o, \quad \text{where}
\end{align*}
\begin{align*}
h_{\tt tag, m_1}(r_o, d_x, v_x, N) = & \sum_{n=0}^{N-1}  \ppmf(N-n-1, \lamr(\pi r_o^2 - {\tt AoI}_2(r_o, d_x, v_x))) \\
& \ppmf(n, \lamr{\tt AoI}_2(r_o, d_x, v_x))  \pcmf(\Ns-n-1, \lamr(\pi r_x^2 - {\tt AoI}_2(r_o, d_x, v_x))).
\end{align*}
In the above expression $r_x(r_o, d_x, v_x) = \sqrt{r_o^2 + d_x^2 - 2 r_o d_x \cos(v_x)}$ and ${\tt AoI}_2(r_o, d_x, v_x)$ is the area of intersection of two circles is given as
\begin{align*}
{\tt AoI}_2(r_o, d_x, v_x) =  r_o^2\left(v_x - \frac{\sin\left(2v_x\right)}{2}\right) + r_x^2 \left(u(r_o, d_x, v_x) - \frac{\sin\left(2u(r_o, d_x, v_x)\right)}{2}\right), \numberthis
\label{eq:AoI_2Circles}
\end{align*}
where 
\begin{align*}
u(r_1, r_2, v) = \arccos\left(\frac{r_2 - r_1 \cos(v)}{\sqrt{r_1^2 + r_2^2 - 2 r_1 r_2 \cos(v)}}\right). \numberthis
\label{eq:Ang_2Circles}
\end{align*}

The corresponding second moment is given as
\begin{align*}
\nbbE[K_N^2] = 2 \pi \lamr \lamu^2 \int\limits_{r_o = 0}^{\infty} \int\limits_{d_x=0}^{\infty} \int\limits_{d_y=0}^{\infty} \int\limits_{v_x=0}^{2 \pi} \int\limits_{v_y=0}^{2 \pi} h_{\tt tag, m_2}(r_o, d_x, d_y, v_x, v_y, N) {\rm d}v_y {\rm d}v_x d_y {\rm d}d_y d_x{\rm d}d_x r_o{\rm d}r_o,
\end{align*}
where $h_{\tt tag, m_2}(r_o, d_x, d_y, v_x, v_y, N)$ is given by \eqref{eq:htagM2} in Appendix~\ref{app:LoadTagRRH}.
\end{lemma}
\begin{IEEEproof}
Please refer to Appendix~\ref{app:LoadTagRRH}.
\end{IEEEproof}

\begin{remark}
The load on a tagged AP depends on its distance from the typical user at the origin. Using the results of the above lemma, we conclude that $\nbbE[K_1] > \nbbE[K_2] > \nbbE[K_3]> \ldots$, and $\nbbE[|K_1|^2] > \nbbE[|K_2|^2] > \nbbE[|K_3|^2]> \ldots$. 
\end{remark}

\subsubsection{Approximation of the load $\pmf$}
With the knowledge of the first two moments of the load, we approximate it as a negative binomial random variable. The $\pmf$ of this random variable with parameters $r$ and $p$ is given as 
\begin{align*}
\nbbP[K_N = k] = \frac{\Gamma(k + r)}{k! \Gamma(r)} p^r (1 - p)^k,
\end{align*}
where $\nbbE[K_N] = (1-p)r/p$ and $\nbbE[K_N^2] = (1-p)r(1 + (1-p)r)/p^2$.
Using the results of Lemma~\ref{lem:LoadTagRRH}, we solve the aforementioned two equations to obtain the values of $r$ and $p$.
The intuition behind the consideration of negative binomial stems from the following fact: if we consider that each user is served by its nearest AP, then the serving region of each AP is its own Poisson-Voronoi cell. The area of this cell is well approximated as a gamma random variable~\cite{ferenc2007size}. Now, conditioned on the area of the cell, the number of users that fall in this area follows Poisson distribution. Hence, once we decondition over the serving area, we get a Poisson-gamma mixture distribution for the number of users served by a tagged AP. Since negative binomial distribution is a consequence of the Poisson-gamma mixture distribution, this justifies our choice to approximate the load with this distribution. 
This approximation will be validated in Sec.~\ref{sec:Results}. 

\subsection{The load of the typical AP}\label{subsec:TypAP}
In this section, we derive the approximate ${\tt PMF}$ of the number of users served by the typical AP in the network. 
Similar to the previous case, since exact characterization of the ${\tt PMF}$ is intractable, we first derive the exact expression for the first two moments of the load for the typical AP. 
Next, using moment-matching method, we approximate the ${\tt PMF}$ as a negative binomial $\pmf$.
The derivation of the first two moments now becomes the special case of the tagged AP result. In the following lemma, we present the first two moments. 
\begin{lemma}\label{lem:LoadTypRRH}
The first two moments of the number of users $K_o$ served by the typical AP are 
\begin{align*}
\nbbE[K_o] = \Ns \frac{\lamu}{\lamr}, \quad \nbbE[K_o^2] = 2 \pi \lamu^2 \int_{r_x=0}^{\infty} \int_{r_y=0}^{\infty} \int_{u=0}^{2 \pi} h_{\tt typ, m2}(r_1, r_2, u) {\rm d}u r_2{\rm d}r_2 r_1{\rm d}r_1 + \Ns \frac{\lamu}{\lamr},
\end{align*}
where $h_{\tt typ, m2}(r_x, r_y, u) =$
\begin{align*}
 \sum_{l=0}^{\Ns-1} & \bigg[\ppmf(l, \lamr {\tt AoI}_2(r_x, r_y, v_{xy}))  \pcmf (\Ns-l-1, \lamr (\pi r_x^2 - {\tt AoI}_2(r_x, r_y, v_{xy})))\\ 
& \times  \pcmf (\Ns-l-1, \lamr (\pi r_y^2 - {\tt AoI}_2(r_x, r_y, v_{xy}))) \bigg]. \numberthis
\label{eq:InnerIntegrand}
\end{align*}
In the above equation, $v_{xy} = \arccos\left(\frac{r_x - r_y \cos(u)}{\sqrt{r_x^2 + r_y^2 - 2 r_x r_y \cos(u)}}\right)$, and ${\tt AoI}_2$ is given in \eqref{eq:AoI_2Circles}. 
\end{lemma}
\begin{IEEEproof}
Please refer to Appendix~\ref{app:LoadTypRRH}.
\end{IEEEproof}

Similar to the tagged AP case, we approximate the load of the typical AP as a negative binomial random variable. 

\begin{remark}
We observe that $\nbbE[K_o^2] \approx \nbbE[K_o]^2 + 1.2802\Ns$. 
This result is exact for $\Ns = 1$~\cite{baccelli1997st}.
\end{remark}

\begin{remark}
The load distribution result of the typical AP also characterizes the degree distribution in an {AB random geometric graph} (AB-RGG). An AB-RGG is a bipartite random graph between two sets of vertices $A$ and $B$, where a point in $A$ is connected to a few points in $B$ based on certain distance criteria~\cite{iyer2012pe, penrose_2014}. In our case, $A= \fir, B = \fiu$ and an edge exists if $\nbx \in \fiu$ is served by $\nby \in \fir$. It is worth mentioning that a simulation-based approximation result for the degree distribution of this type of AB-RGG is recently proposed in~\cite{stegehuis2021de}.
\end{remark}

\section{Rate Coverage for user-centric cell-free mMIMO}\label{sec:RateUC}
The achievable rate result derived in Lemma~\ref{lem:RoLB} is directly applicable to the user-centric cell-free architecture. Recall that $\fir$ and $\fiu$ are two independent homogeneous PPP. Further, we consider that the typical user at the origin and its set of serving APs $\Phi_{ro}$ consists of the nearest $\Ns$ APs. 
Determining the distribution of the rate or $\sinr$ of the typical user is intractable as each term in \eqref{eq:SINR_LB_Gen} depends on a set of common distances. However, a degree of tractability can be achieved for theoretical analysis by assuming that the network is operating in a regime where pilot contamination is negligible. 
There are two consequences of this assumption. First, we can ignore the pilot contamination interference term in \eqref{eq:SINR_LB_Gen}. 
Second, the variance of the channel estimate presented in~\eqref{eq:EstVariance} can be approximated as 
\begin{align*}
\gamma_{ok} = \frac{\tp \rp \beta_{ok}^2}{1 + \tp \rp \sum_{\nbu_j \in \Phi_{u}} \boldsymbol{\psi}(k)^H \boldsymbol{\psi}(j) \beta_{oj}} \approx \frac{\tp \rp \beta_{ok}^2}{1 + \tp \rp \beta_{ok}}.
\end{align*}
\begin{remark}
A system with a pilot allocation scheme that ensures that each AP serves only one user per pilot is likely to operate in the regime where the above assumption holds, especially, for the set of serving (dominant) APs. This pilot assignment is realizable in a low mobility scenario, where the coherence block is sufficiently large. For example, if we consider a coherence bandwith of 200 kHz and a coherence time of 2 ms, then the TDD coherence block has 400 symbols. Let us assume that each AP is served by the nearest $\Ns = 5$ APs. In such a scenario, the probability that the set of tagged APs, which serve the typical user, collectively serve more than 30 users is less than 0.002. Hence, the above criteria is met by reserving around 30 symbols, which is less than 10\% of the coherence block, for pilot estimation.
\end{remark}

With this assumption, we present the rate coverage for the typical user. 
\begin{prop}\label{prop:RateCov}
The rate coverage for the typical user is given as $\Rcinf = \nbbP[\se_o > \Tr] = $
\begin{align*}
& \int\limits_{d_{o\Ns}= 0}^{\infty} \int\limits_{d_{o\Ns-1}= 0}^{d_{o\Ns}} \ldots \int\limits_{d_{o1}= 0}^{d_{o\Ns}} \sum_{k_1 = 0}^{\infty} \mathbf{1}\bigg(2 \pi \lamr \int\limits_{d_{o\Ns}}^{\infty} l(r) r{\rm d}r \leq h_{\tt cov}(k_1, d_{o1}, d_{o2}, \ldots, d_{o\Ns}) \bigg) \\
& \times \nbbP[K_1 = k_1] \prod_{i=1}^{\Ns -1} \frac{2 d_{oi}}{d_{o\Ns}} f_{D_{o\Ns}}(d_{o\Ns}) {\rm d}d_{o1} \ldots {\rm d}d_{o\Ns},
\end{align*}
where $ h_{\tt cov}\left(k_1, d_{o1}, d_{o2}, \ldots, d_{o\Ns} \right) =$
\begin{align*}
 & \frac{\Na}{(2^{\Tr}-1)K_{\tt max}} \left(\sqrt{\gamma_{o1}(1 - 2^{-\frac{\Cf}{\min\{k_1+1, K_{\tt max}\}}})} + \sum_{l=2}^{\Ns} \sqrt{\gamma_{ol}(1 - 2^{-\Cf/\bar{K}_l})}\right) \\ 
 & - \frac{\Na}{K_{\tt max}}\left(\gamma_{o1}2^{-\frac{\Cf}{\min\{k_1+1, K_{\tt max}\}}} + \sum_{l=2}^{\Ns} \gamma_{ol}2^{-\Cf/\bar{K}_l}\right) - \sum_{l=1}^{\Ns} l(d_{ol}) - \frac{1}{\rd}.\numberthis
 \label{eq:RHSofRateCov}
\end{align*}
In the above expression, $d_{o1} \leq d_{o2} \leq \ldots \leq d_{o\Ns}$, where $d_{oi} $ is the distance between the typical user and its $i$-th closest AP. The load of the closest AP is $K_1$ and the mean load of the $i$-th closest AP is $\bar{K}_i$, which is given as $ \bar{K}_i = 1 + \sum_{k_i = 0}^{\infty} \min\{k_i, K_{\tt max}\} \nbbP[K_i = k_i]$.
\end{prop}
\begin{IEEEproof}
Please refer to Appendix~\ref{app:RateCov}. 
\end{IEEEproof}

\section{Results and Discussion}\label{sec:Results}
In this section, we provide useful system design insights from our analysis as well as validate our theoretical rate coverage results through extensive Monte Carlo simulations. 

\subsection{Performance of traditional cell-free mMIMO}
First, we validate the theoretical results derived for the traditional architecture in Sec.~\ref{sec:RateTrad}. 
We have considered a finite circular region of radius $\Rs = 500$ m. The path loss function is
\begin{align*}
l(r) = & r^{3.7}\mathbf{1}(r>1) + \mathbf{1}(r\leq 1). \numberthis
\label{eq:PathLossFun}
\end{align*}
We consider the transmit $\snr$ $\rd = 100$ dB\footnote{The received $\snr$ at the edge of the system from the center with the considered path loss model is 0.1381 dB.} and transmit pilot $\snr$ $\rp = 100$ dB. Let the TDD coherence block consists of 400 symbols, which corresponds to a coherence bandwidth of 200 kHz and a coherence time of 2 ms. Further, the pilot sequence is of length $\tp = 80$ symbols unless stated otherwise. We assume that the users are assigned orthogonal pilot sequences, equivalently the number of users in this finite system $K \leq \tp$.
Note that as stated before Lemma~\ref{lem:RoLB}, we only present the user or system $\se$ (in terms of bits/s/Hz). These values will scale depending on the system bandwidth and the fraction of coherence block dedicated for the downlink transmission.
The choice of other system parameters are indicated at necessary places.
Using the rate coverage result of Proposition~\ref{prop:RcFinite}, the average user rate is expressed as 
\begin{align*}
\bar{\se}_{o,{\tt fin}} = \int_{t=o}^\infty t \mathtt{R_{c, fin}}(t) {\rm d}t \quad \text{bits/s/Hz},
\end{align*}
and corresponding average system sum-rate is $K \bar{\se}_{o,{\tt fin}}$ {bits/s/Hz}.

\subsubsection{The effect of fronthaul capacity} 
\begin{figure*}[!htb]
\centering
\begin{subfigure}{0.35\textwidth}
  \centering
  \includegraphics[width=\linewidth]{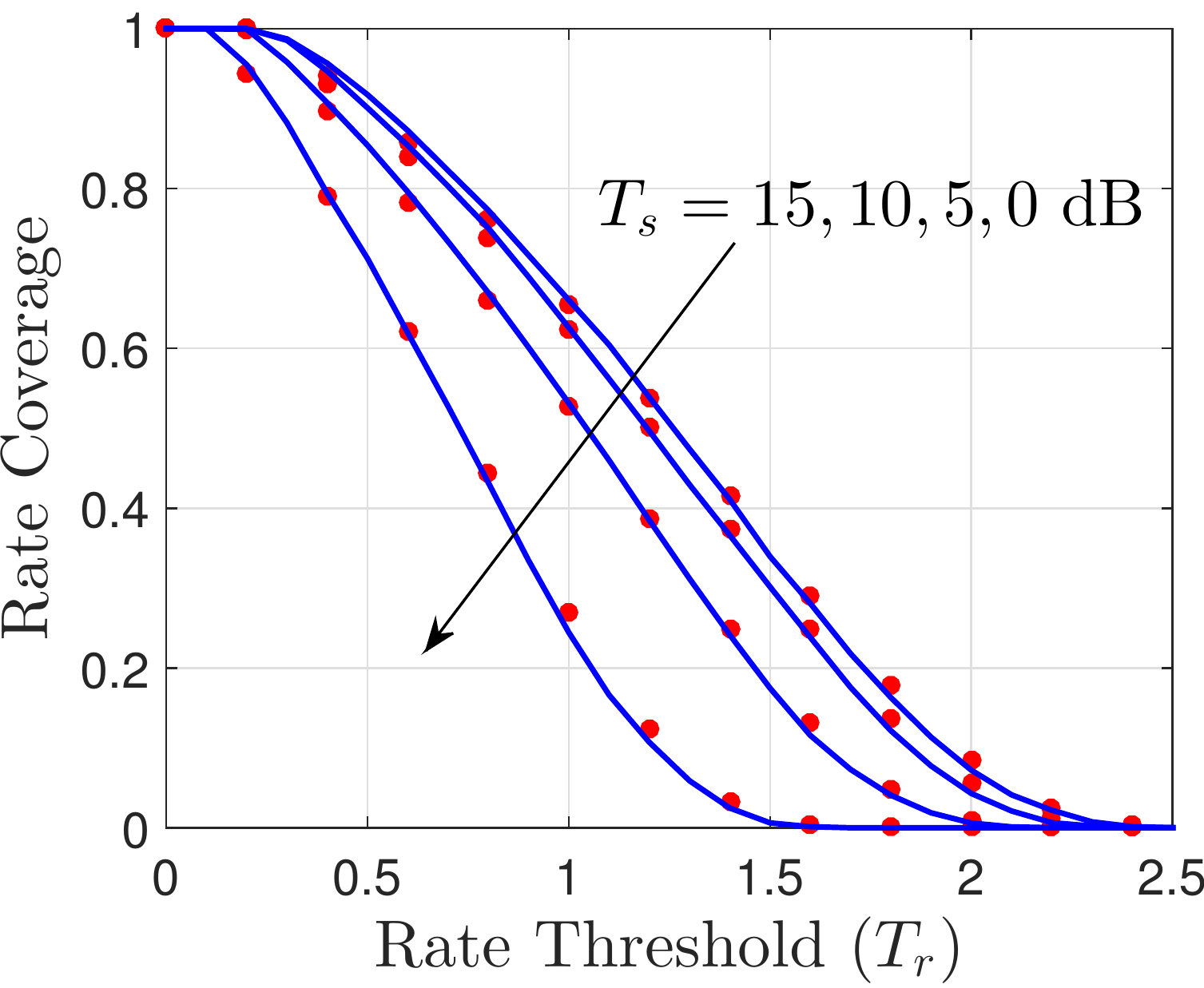}
\end{subfigure}
\hspace{0.4cm}
\begin{subfigure}{0.35\textwidth}
  \centering
  \includegraphics[width=\linewidth]{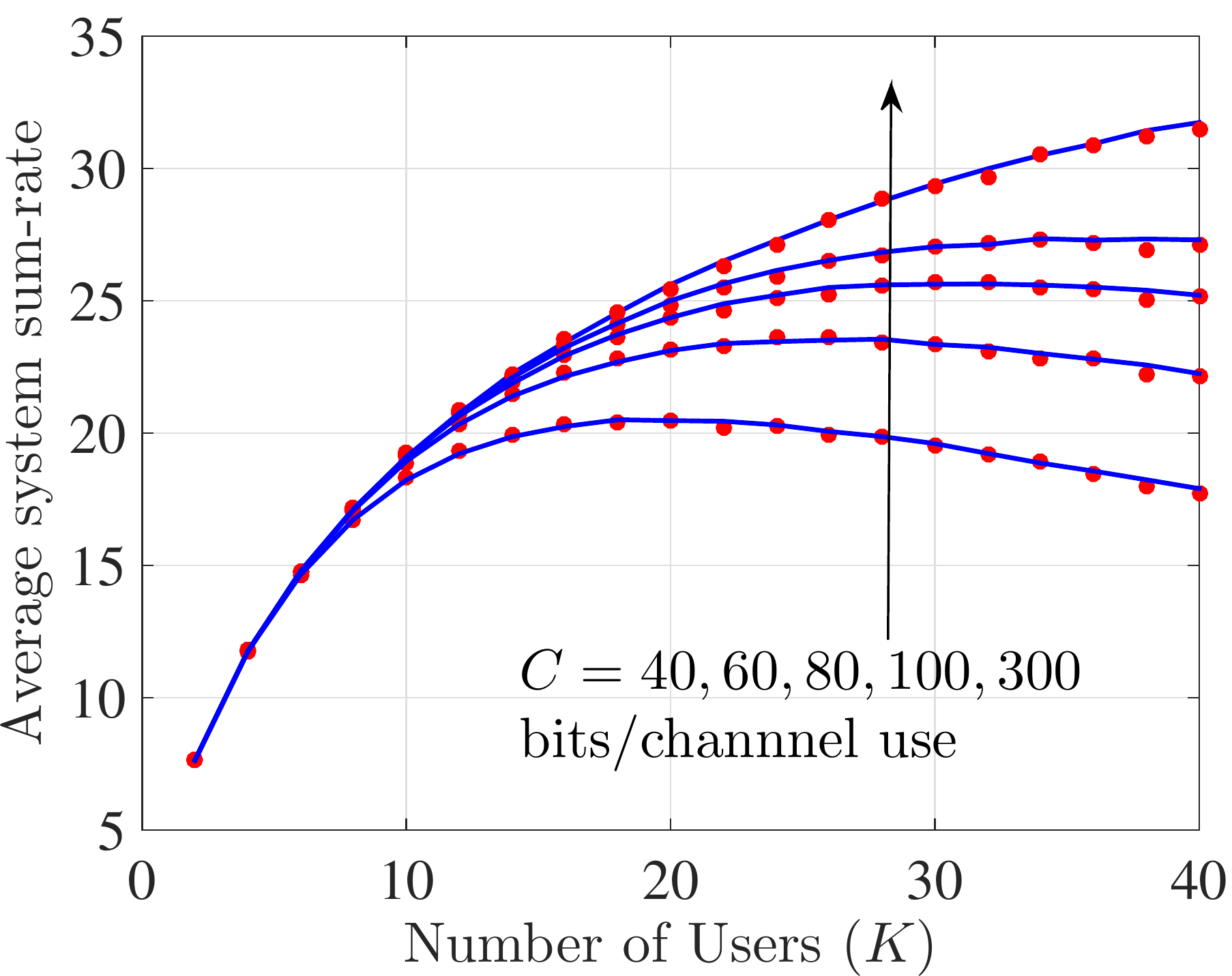}
\end{subfigure}%
\caption{\small The effect of fronthaul capacity on system performance. The solid lines are obtained using the analytical results presented in Sec~\ref{sec:RateTrad} and markers are Monte Carlo simulation results. The system parameters are $M =32, \Na= 4, \tp = 80, \rp = \rd = 100$ dB.}
\label{fig:CSE_Vs_K_For_BakC}
\end{figure*}

In Fig.~\ref{fig:CSE_Vs_K_For_BakC} (left), we have presented the rate coverage  of the system for $K = 20$ users in the system. As expected, the rate coverage improves with increasing ${\tt SCNR}$ threshold $\Ts$ as it directly corresponds to a higher fronthaul capacity $\Cf$. Further, in Fig.~\ref{fig:CSE_Vs_K_For_BakC} (right), the average system sum-rate is presented as a function of the number of users $K$ for different fronthaul capacities.
We have kept a high pilot transmission $\snr$ $\rp = 100$ dB corresponding to an almost perfect CSI scenario to highlight the effect of fronthaul capacity on the system performance.
As evident from the figure, the average system sum-rate is quasi-concave function of the number of users. Further, for a given number of APs, the optimum number of users that should be multiplexed to maximize the average rate increases with the increasing fronthaul capacity. 
From the trend, we infer that when $\Cf$ has unlimited capacity, the maximum average system sum-rate is obtained by serving all the users simultaneously.

\subsubsection{Distributed vs. collocated}\label{subSec:distColloc}
\begin{figure*}[!htb]
\centering
\begin{subfigure}{0.35\textwidth}
  \centering
  \includegraphics[width=\linewidth]{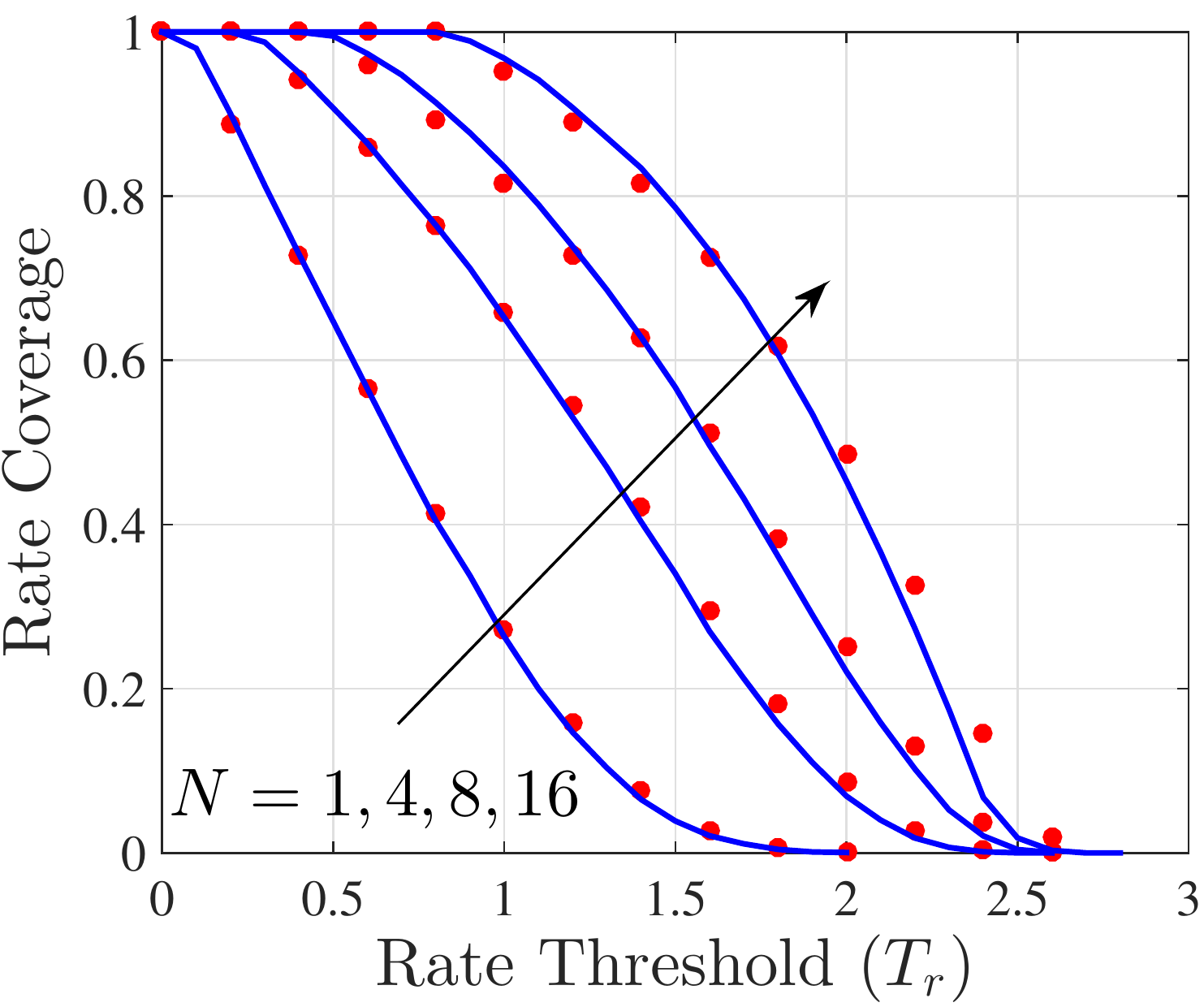}
\end{subfigure}
\hspace{0.4cm}
\begin{subfigure}{0.35\textwidth}
  \centering
  \includegraphics[width=\linewidth]{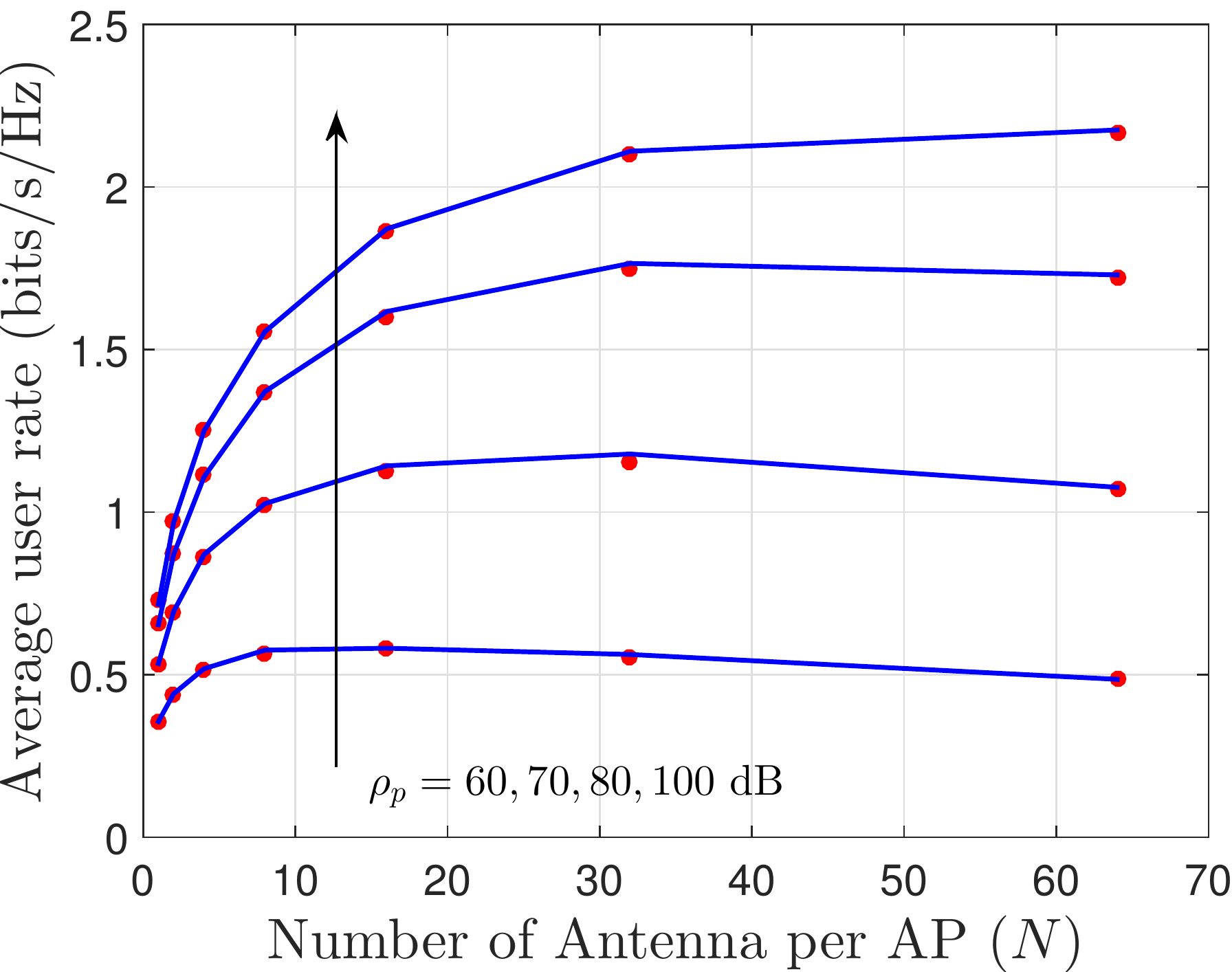}
\end{subfigure}%
\caption{\small The effect of number of antennas per AP on rate coverage (left) and average user rate (right). Solid lines and markers represent analytical and simulation results, respectively. We have considered $M\Na= 128, \tp = 80, K = 20, \rd = 100 \ \text{dB},{\tt SCNR} = 15$ dB.}
\label{fig:CSE_Vs_N_For_K}
\end{figure*}

In Fig.~\ref{fig:CSE_Vs_N_For_K} (left), we present the rate coverage of the system for different number of antennas at the AP while keeping the total number of antennas in the system fixed, i.e., $M\Na= 128$. 
We observe that the centralized architecture performs better than a distributed architecture under conjugate beamforming and equal power allocation. 
In Fig.~\ref{fig:CSE_Vs_N_For_K} (right), we present the average user rate for different number of antennas at each AP while keeping the total number of antennas in the service region fixed. We consider the ${\tt SCNR}$ threshold $\Ts = 15$ dB. We observe that for high $\rp$ (i.e., high-quality CSI), average user rate increases as we move towards a more collocated setup. 
On the other hand, with low $\rp$ (i.e., low-quality CSI), the average user $\se$ is a concave function of the number of antennas per AP. This behavior is in contrast to the conventional MIMO results where a distributed implementation is always preferable. The justification to this counter-intuitive trend can be explained by the fact that due to conjugate beamforming, we get a self-interference term from all the APs as evident from the $\sinr$ expression in \eqref{eq:SINR_LB_JS}. Hence, with a distributed implementation, the desired signal power from the nearest AP increases and so does the self interference term. Therefore, a more collated set up is preferable.

\subsection{Performance of user-centric cell-free mMIMO}
Now, we verify the theoretical results derived in Secs.~\ref{sec:LoadDerivation} and \ref{sec:RateUC} through extensive Monte Carlo simulations. Further, we provide a few useful network dimensioning guidelines. 
For the simulations, we consider the system radius to be 2000 m. The path loss is the same as \eqref{eq:PathLossFun}. 
Further, we consider $\rd = \rp = 100$ dB and $\tp = 80$.
The choice of other system parameters are indicated at appropriate places.

\subsubsection{The load distribution result}
\begin{figure*}[!htb]
\centering
\begin{subfigure}{0.4\textwidth}
  \centering
  \includegraphics[width=\linewidth]{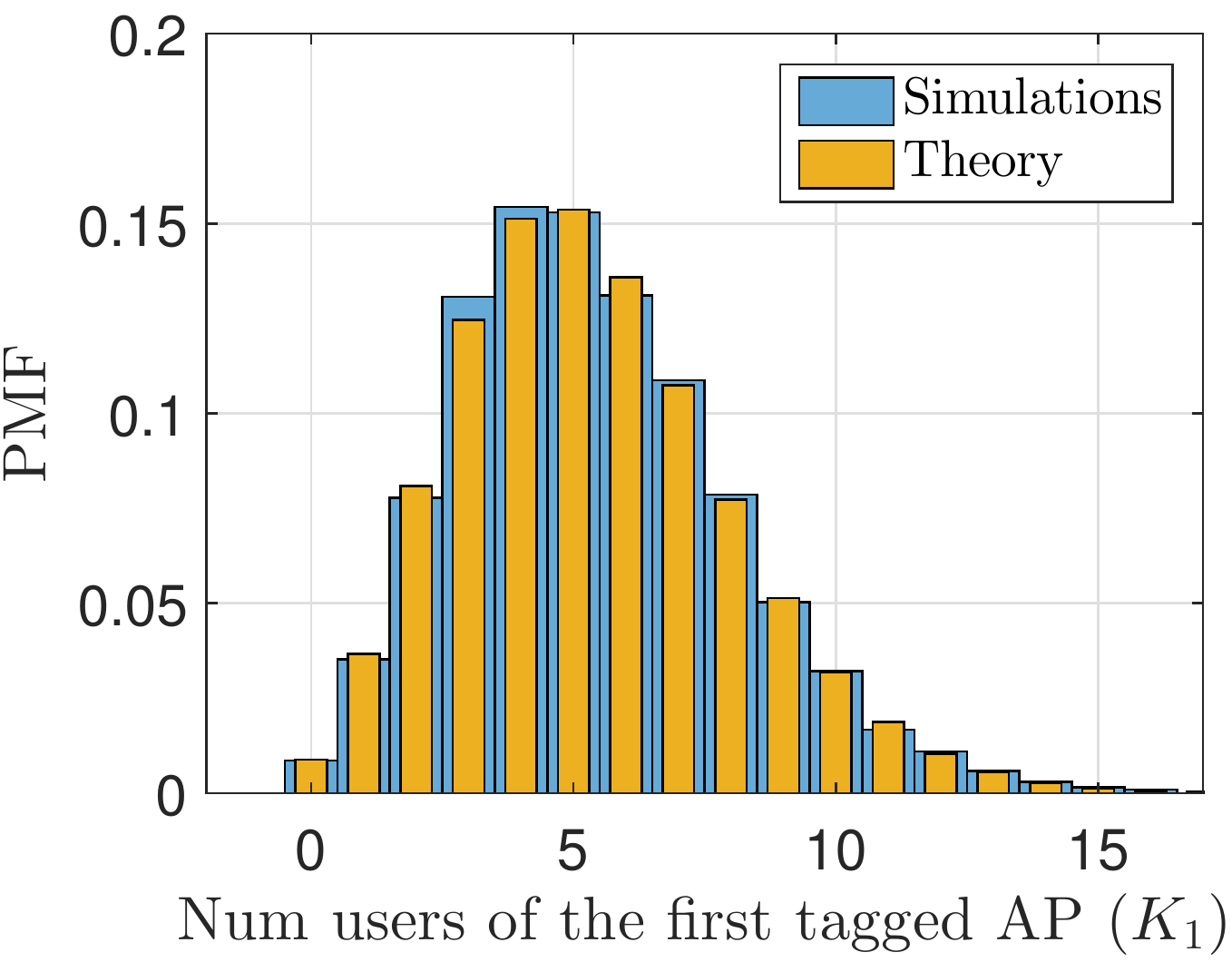}
\end{subfigure}
\hspace{0.3cm}
\begin{subfigure}{0.4\textwidth}
  \centering
  \includegraphics[width=\linewidth]{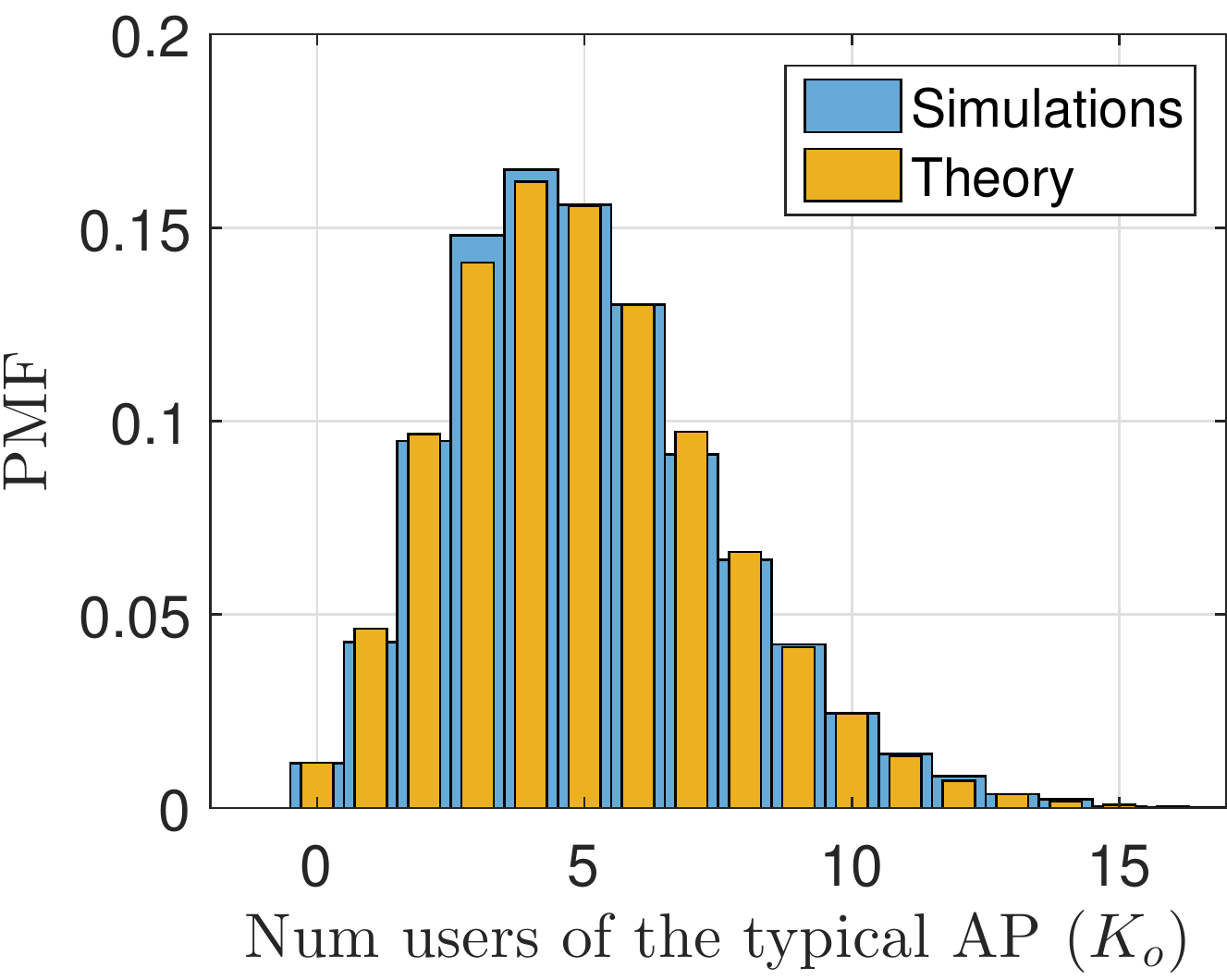}
\end{subfigure}%
\caption{\small The $\pmf$ of the number of users served by the nearest tagged AP (left) and the typical AP (right) in the network. $\lamr = \lamu = 10^{-4}, \Ns = 5$. }
\label{fig:APLoad}
\end{figure*}

In Fig.~\ref{fig:APLoad}, we validate the approximate theoretical $\pmf$s derived in Secs.~\ref{subsec:TagAP} and \ref{subsec:TypAP}, respectively, by Monte Carlo simulations. In both the cases, theoretical  and simulation results are remarkably close.
In Fig.~\ref{fig:LoadIns} (left), using the typical AP load distribution result, for a given fronthaul capacity $\Cf$ and ${\tt SCNR}$ threshold $\Ts$, we present the probability that ${\tt SCNR}$ is above $\Ts$ as a function of number of serving APs $\Ns$. 
Formally, we study 
\begin{align*}
\nbbP[{\tt SCNR} \geq \Ts] = \nbbP[2^{\Cf/K_o} - 1\geq \Ts] = \nbbP[K_o \leq \Cf/\log_2(1 + \Ts)],
\end{align*}
which is the $\cdf$ of the typical AP load. 
As expected, the more stringent the $\Ts$, the lower the probability of having a ${\tt SCNR}$ more than $\Ts$ for a given number of serving APs. Note that in this result we assume that all the users attached to the typical AP are scheduled on the same resource while using different pilots. 
In Fig.~\ref{fig:LoadIns} (right), we present the required fronthaul capacity as a function of number of serving APs. As expected, we observe a linear growth in $\Cf$ with increasing $\Ns$. However, the rate of growth depends on how strict the ${\tt SCNR}$ constraint is. 

\subsubsection{The rate coverage result and insights}

\begin{figure*}[!htb]
\centering
\begin{subfigure}{0.35\textwidth}
  \centering
  \includegraphics[width=\linewidth]{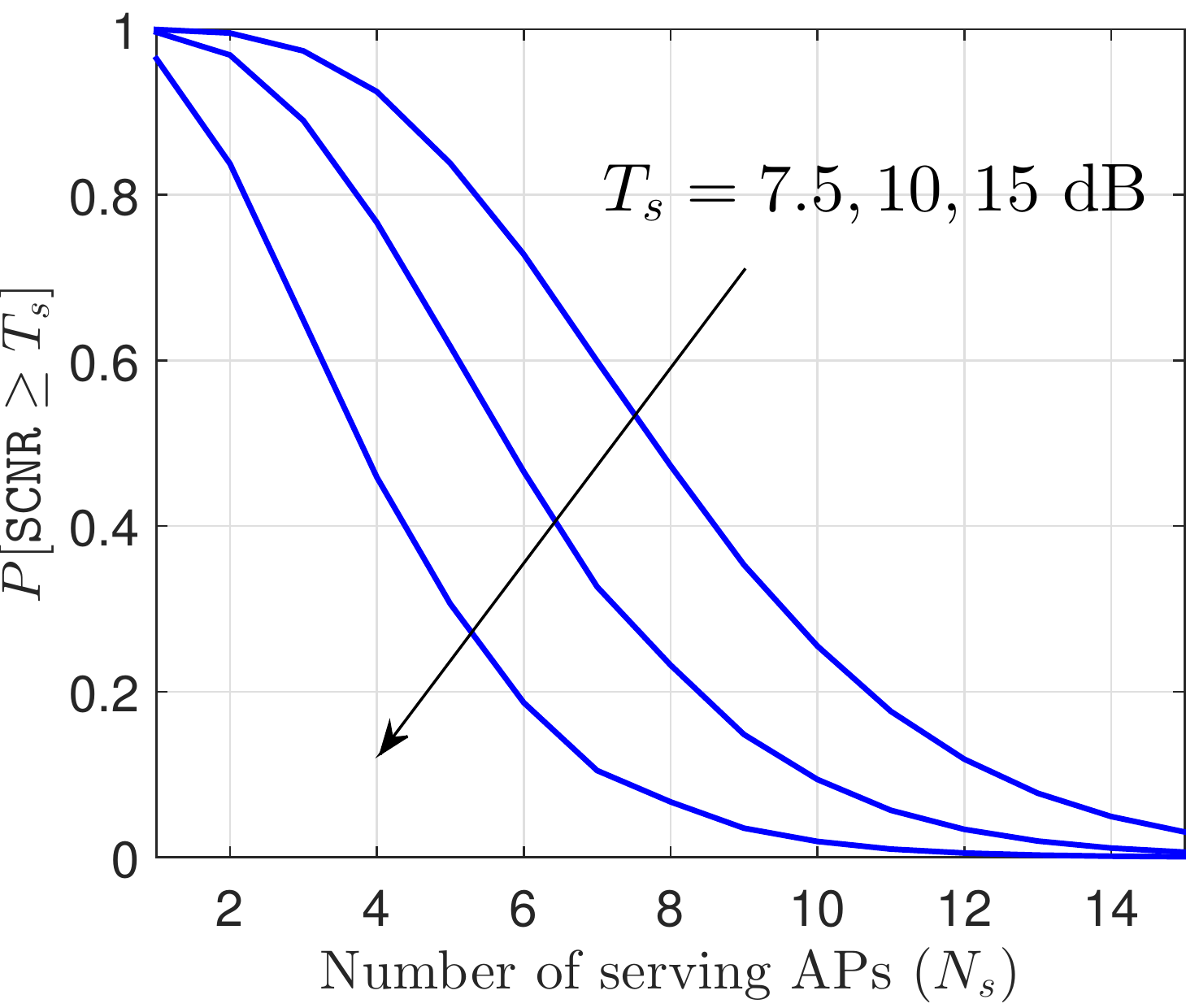}
\end{subfigure}
\hspace{0.4cm}
\begin{subfigure}{0.35\textwidth}
  \centering
  \includegraphics[width=\linewidth]{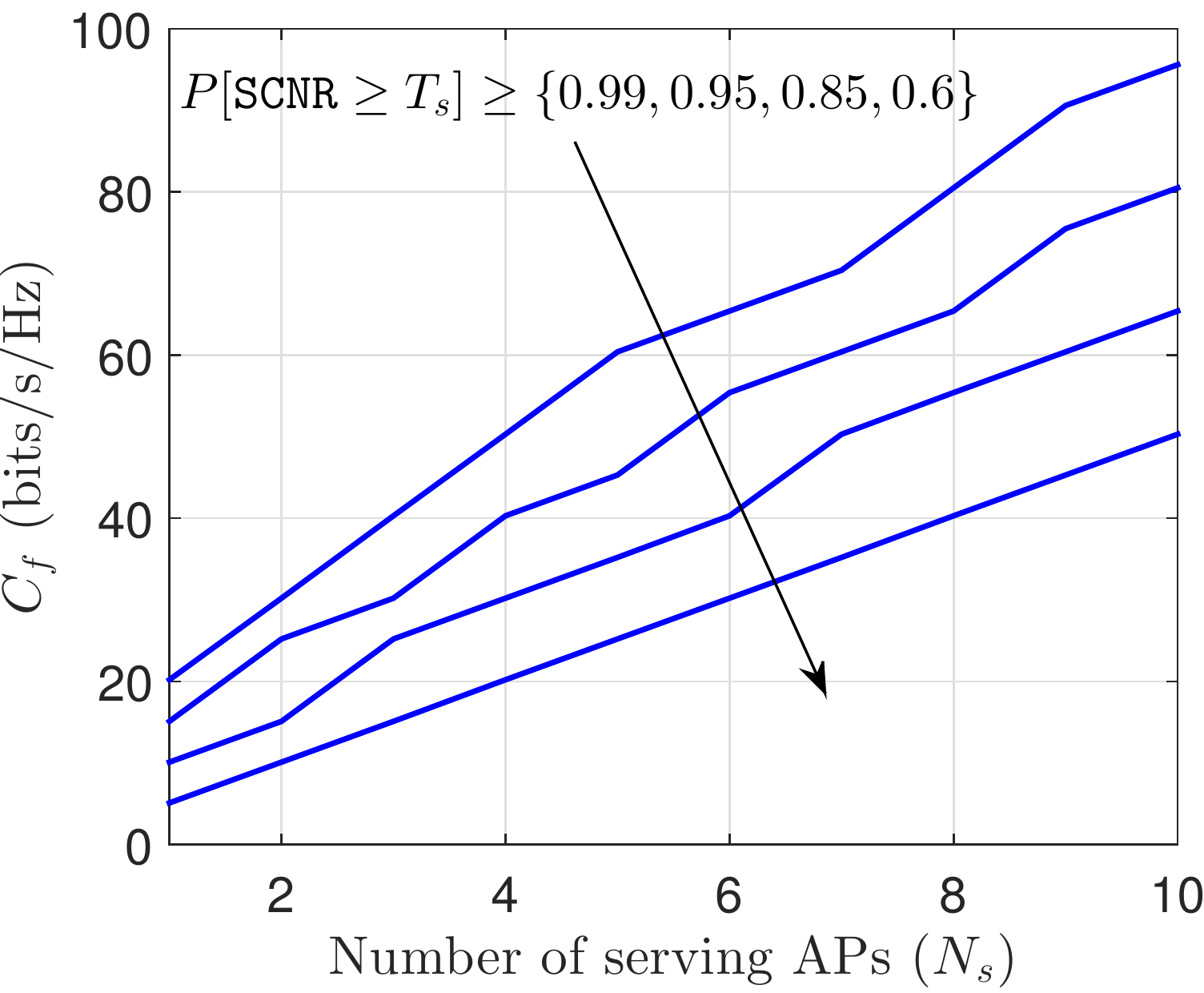}
\end{subfigure}%
\caption{\small (Left) The number of serving APs to ensure a certain minimum ${\tt SCNR}$ for different $\Ts$. Other parameters: $\lamu = \lamr = 10^{-4}, \Cf = 20$ bits/s/Hz. (Right) The required fronthaul capacity as a function of $\Ns$ to ensure $\nbbP[{\tt SCNR} \geq \Ts]$ is above a certain threshold. Other parameters: $\lamu = \lamr = 10^{-4}, \Ts = 15$ dB.} 
\label{fig:LoadIns}
\end{figure*}

\begin{figure*}[!htb]
\centering
\begin{subfigure}{0.35\textwidth}
  \centering
  \includegraphics[width=\linewidth]{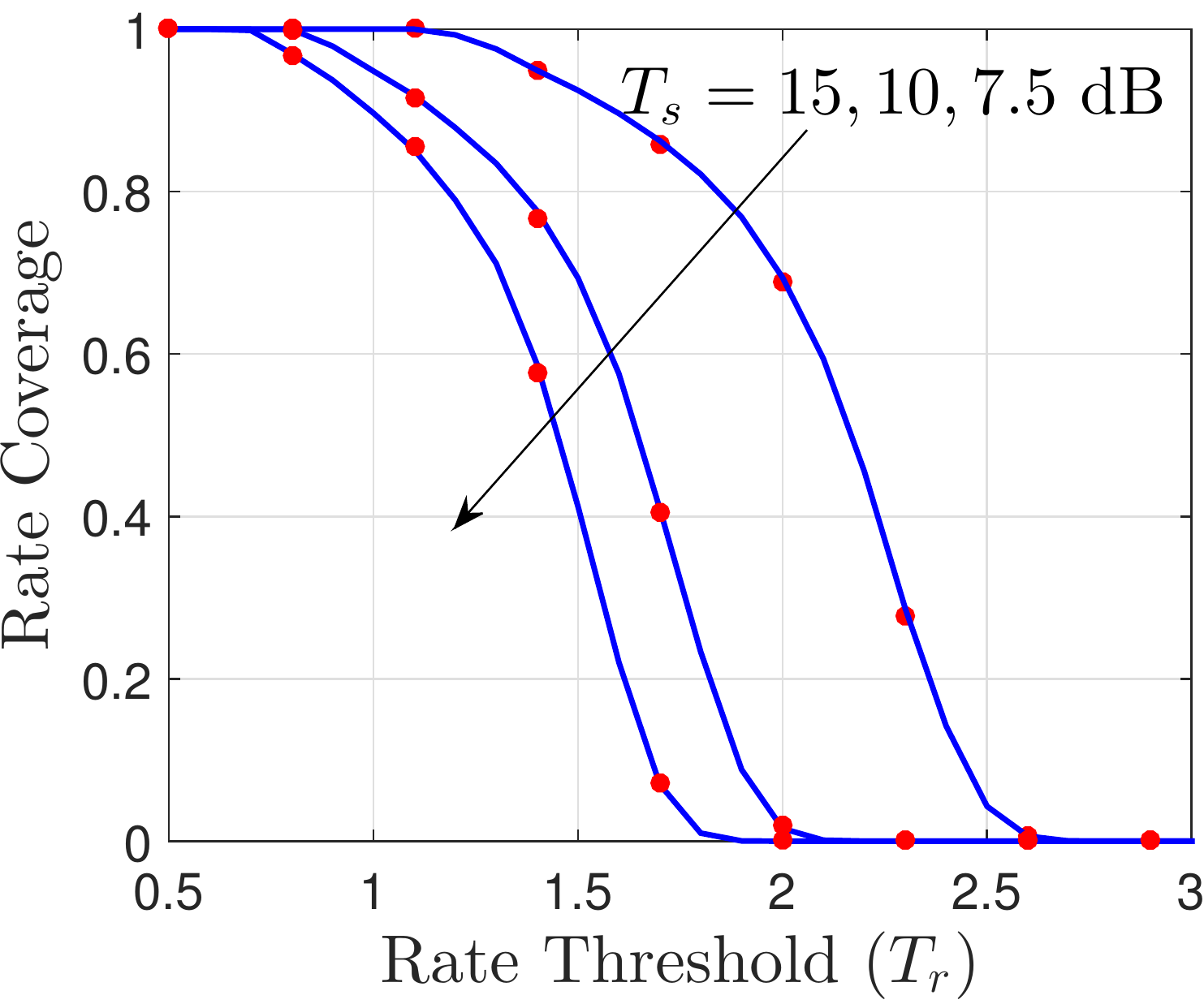}
\end{subfigure}
\hspace{0.4cm}
\begin{subfigure}{0.35\textwidth}
  \centering
  \includegraphics[width=\linewidth]{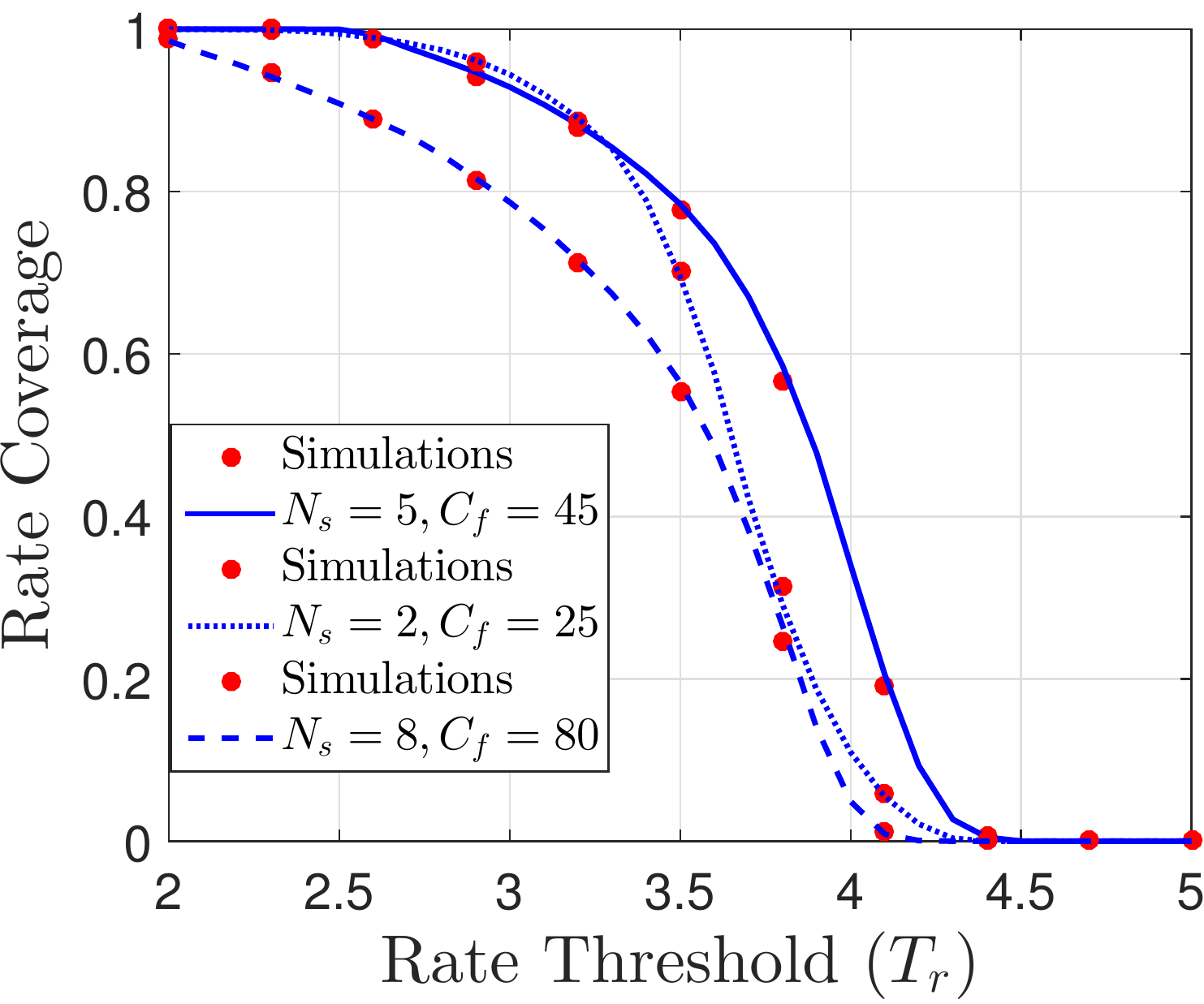}
\end{subfigure}%
\caption{\small (Left) The rate coverage of the typical user for different $\Ts$ and a  fixed $\Cf = 45$ bits/s/Hz. Solid lines and markers represent theory and simulations, respectively. Other system parameters:$\lamu = \lamr = 10^{-4}, \Na= 10, \Ns = 5, \tp = 80, \rp = \rd = 100$ dB. (Right) The rate coverage for different number of serving APs ($\Ns$). Other system parameters: $\lamu = \lamr = 10^{-4}, \Na= 40, \tp = 120, \rp = \rd = 100, \Ts = 15$ dB.}
\label{fig:RateCov}
\end{figure*}

In Fig.~\ref{fig:RateCov} (left), we present the rate coverage for the typical user for different ${\tt SCNR}$ thresholds when it gets scheduled. As observed from the figure, for a fixed $\Cf$, the rate coverage improves with increasing $\Ts$. Note that as per Remark~\ref{rem:MaxLoad}, for a fixed $\Cf$, maximum number of scheduled users per resource unit increases with decreasing $\Ts$. As a result, the power is equally divided among more users that results in a lower coverage. 
In Fig.~\ref{fig:RateCov} (right), we plot the rate coverage of the typical user for different $\Ns$. The corresponding $\Cf$ is selected using the result of Fig.~\ref{fig:LoadIns} such that $\nbbP[{\tt SCNR} \geq \Ts] \approx 0.95$. From the figure, 
we observe that serving a user by more number of APs is not always advantageous. Hence, system should operate at the optimum $\Ns$ for a given set of parameters.

\section{Conclusion}
In this work, we modeled and analyzed a cell-free mMIMO network with finite fronthaul capacity using tools from stochastic geometry.
We considered two different architectures of cell-free mMIMO, namely, the traditional architecture where each AP serves all the users in the network and the user-centric architecture where the typical user is served by a few nearest APs. 
For the traditional architecture, we provided the user rate coverage result using the relevant statistics of BPPs. 
For the user-centric architecture, we characterized the load distribution result for the typical AP as well as the set of tagged APs that serve the typical user in the network.
Further, using the statistical properties of PPPs, we presented the rate coverage result for the typical user in the network. 
From our analyses, we conclude that for the traditional architecture, a more collocated implementation of cell-free mMIMO is preferred over the fully distributed implementation, in the presence of high-quality CSI at the APs. 
Further, for the user-centric architecture, increasing the number of serving APs does not necessarily improve the rate coverage. Hence, there exists an optimal number of serving APs that depends on different system parameters. 
Promising future extensions of this work include the characterization of the scheduling probability of the typical user for the user-centric architecture. This requires knowledge of the total number of users served by the set of tagged APs. 
Further, the system behavior in the presence of pilot contamination is also a promising direction that can lead to useful guidelines for pilot allocation.

\appendix
\subsection{Proof of Lemma~\ref{lem:RoLB}}\label{app:LB}
The proof of this lemma is based on a lower bound that is well known in the mMIMO literature (cf. \cite[Lemma~2]{Caire2017a}, \cite{Medard2000}). 
From~\eqref{eq:ro}, we write $y_o^{\tt dl}$ as 
{\small
\begin{align*}
y_o^{\tt dl} = & \underbrace{\sqrt{\rd} \sum_{\nbr_l \in \Phi_{ro}}  \sqrt{\eta_{lo}} \frac{\dE{\|\hat{\nbg}_{lo}\|^2}{}}{\sqrt{\Na\gamma_{lo}}} q_{o_l}}_{\text{$T_1$: Desired signal}} + \underbrace{\sqrt{\rd} \sum_{\nbr_l \in \Phi_{ro}} \sqrt{\eta_{lo}} \frac{\left(\|\hat{\nbg}_{lo}\|^2 - \dE{\|\hat{\nbg}_{lo}\|^2}{}\right)}{\sqrt{\Na\gamma_{lo}}} q_{o_l}}_{\text{$T_2$: Beamforming uncertainity}}  + \underbrace{\sum_{\nbr_l \in \Phi_{ro}} \frac{\|\hat{\nbg}_{lo}\|^2}{\sqrt{\Na\gamma_{lo}}} \sqrt{\rd \eta_{lo}} \tilde{q}_{o_l}}_{\text{$T_3$: compression error}}  \\ 
& + \underbrace{\sum_{\nbr_l \in \Phi_{ro}} \sqrt{\rd \eta_{lo}} \frac{\tilde{\nbg}_{lo}^T\hat{\nbg}_{lo}^*}{\sqrt{\Na\gamma_{lo}}} \hat{q}_{o_l}}_{\text{$T_4$: chanenel estimation error}} + \underbrace{\sum_{\nbr_l \in \Phi_{ro}} \sum_{\nbu_i \in \Psi_{ul} \setminus \{\nbu_o\}} \sqrt{\rd \eta_{li}} \frac{\nbg_{lo}^T \hat{\nbg}_{li}^*}{\sqrt{\Na\gamma_{li}}} \hat{q}_{i_l}}_{\text{$T_5$: inter user interference}} + \underbrace{\sum_{\nbr_l \in \Phi_{ro}^C} \sum_{\nbu_i \in \Psi_{ul} \setminus \ncalP_o} \sqrt{\rd \eta_{li}} \frac{\nbg_{lo}^T \hat{\nbg}_{li}^*}{\sqrt{\Na\gamma_{li}}} \hat{q}_{i_l}}_{\text{$T_6$: interfering AP signal}} \\
& + \underbrace{\sum_{\nbu_i \in \{\ncalP_o \setminus \{\nbu_o\}\}}  \sum_{\nbr_l \in \Phi_{ri}} \sqrt{\rd \eta_{li}} \frac{\nbg_{lo}^T \hat{\nbg}_{li}^*}{\sqrt{\Na\gamma_{li}}} \hat{q}_{i_l}}_{\text{$T_7$: pilot contamination}} + n_o, \numberthis
\label{eq:app1}
\end{align*}}
$\!$where we assume that the user $\nbu_o$ has the knowledge of the average channel statistics with respect to its serving APs and $\ncalP_o$ contains the user locations that use the same pilot sequence as $\nbu_o$.
From \eqref{eq:app1}, it can be shown that the desired signal is uncorrelated to the rest of the signals. To derive an achievable rate (lower bound on the capacity), we replace the uncorrelated signals by independent Gaussian random variables~\cite{Medard2000}. The variance of each of the Gaussian variables is equal to the corresponding variance of the undesired signal. Hence, the $\sinr$ corresponding to this achievable rate is given as 
$
\sinr_o = {\dE{|T_1|^2}{}}/{(\sum_{i=2}^7 \dE{|T_i|^2}{} +1)}.
$
In this case, note that $\dE{T_i}{} = 0$ for all $i$. Further,
{\small
\begin{align*}
\dE{|T_1|^2}{} & = \rd \Na\left(\sum\limits_{\nbr_l \in \Phi_{ro}}  \sqrt{\frac{\gamma_{lo} (1 - 2^{-\Cf/k_l})}{K_{\tt max}}}\right)^2, \dE{|T_2|^2}{} =  \rd \sum\limits_{\nbr_l \in \Phi_{ro}}  \frac{\gamma_{lo} (1-2^{-\Cf/k_l})}{K_{\tt max}},\\
\dE{|T_3|^2}{} & = \rd (\Na+1) \sum\limits_{\nbr_l \in \Phi_{ro}} \frac{\gamma_{lo}}{K_{\tt max}} 2^{-\Cf/k_l}, \dE{|T_4|^2}{} = \rd \sum\limits_{\nbr_l \in \Phi_{ro}} \frac{(\beta_{lo} - \gamma_{lo})}{K_{\tt max}}, \\
\dE{|T_5|^2}{} & = \rd \sum\limits_{\nbr_l \in \Phi_{ro}} \frac{k_l - 1}{K_{\tt max}} \beta_{lo} \leq  \rd \sum\limits_{\nbr_l \in \Phi_{ro}} \frac{K_{\tt max} - 1}{K_{\tt max}} \beta_{lo} , \quad \dE{|T_6|^2}{} = \rd \sum\limits_{\nbr_l \in \Phi_{ro}^c} \leq \frac{K_{\tt max} - 1}{K_{\tt max}} \beta_{lo}, \\
\dE{|T_7|^2}{} & = \rd \sum\limits_{\nbu_i \in \{\ncalP_o \setminus \nbu_o\}} \left(\sum\limits_{\nbr_l \in \Phi_{ri}} \frac{\beta_{lo}}{K_{\tt max}} + \Na\left(\sum\limits_{\nbr_l \in \Phi_{ri}} \sqrt{\frac{\gamma_{lo}}{K_{\tt max}}}\right)^2\right).
\end{align*}
}
Substituting these values, we obtain the expression presented in the lemma. \hfill \IEEEQED

\subsection{Proof of Lemma~\ref{lem:LoadTagRRH}}\label{app:LoadTagRRH}
\begin{figure*}[!htb]
\centering
\begin{subfigure}{0.3\textwidth}
  \centering
  \includegraphics[width=\linewidth]{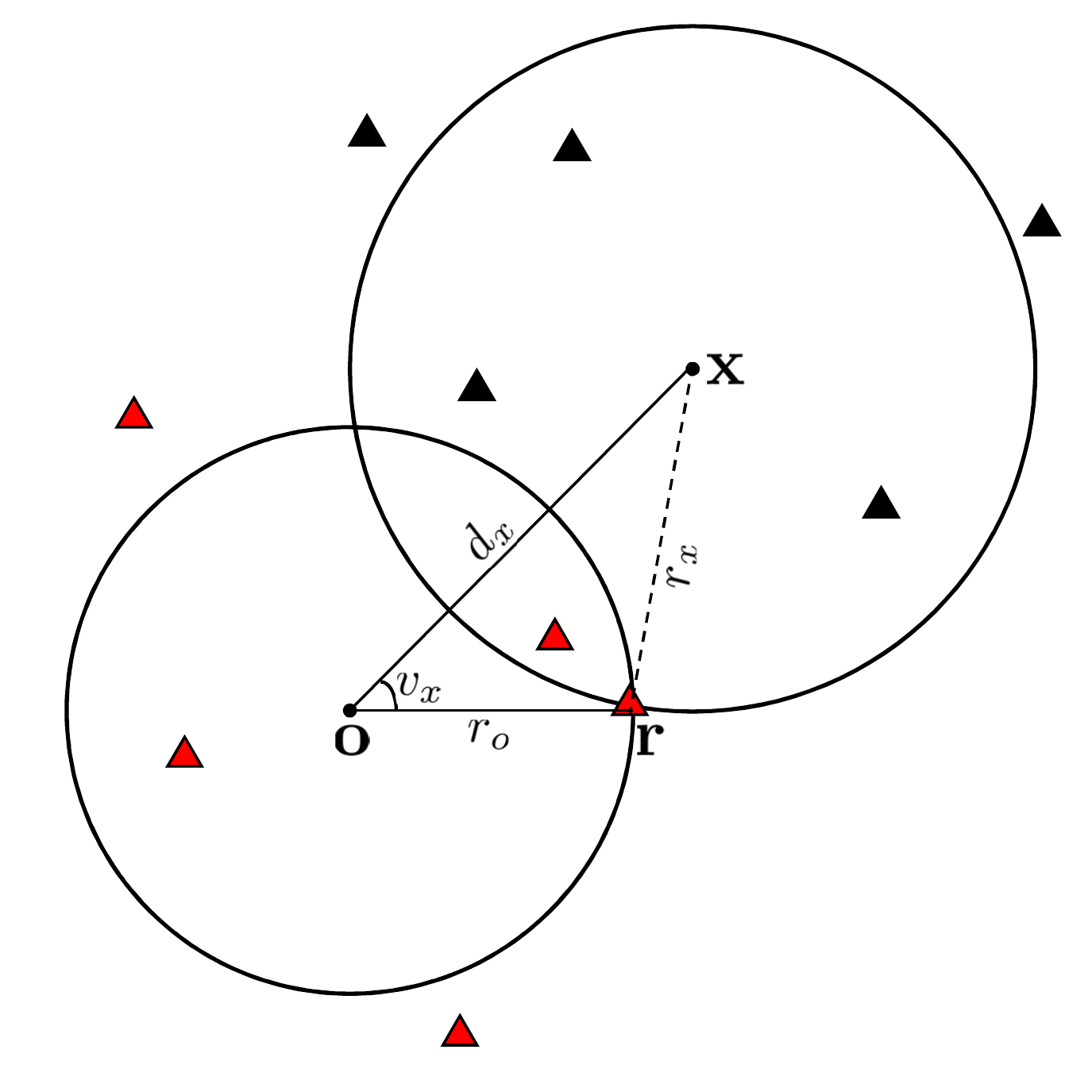}
\end{subfigure}
\hspace{0.5cm}
\begin{subfigure}{0.3\textwidth}
  \centering
  \includegraphics[width=\linewidth]{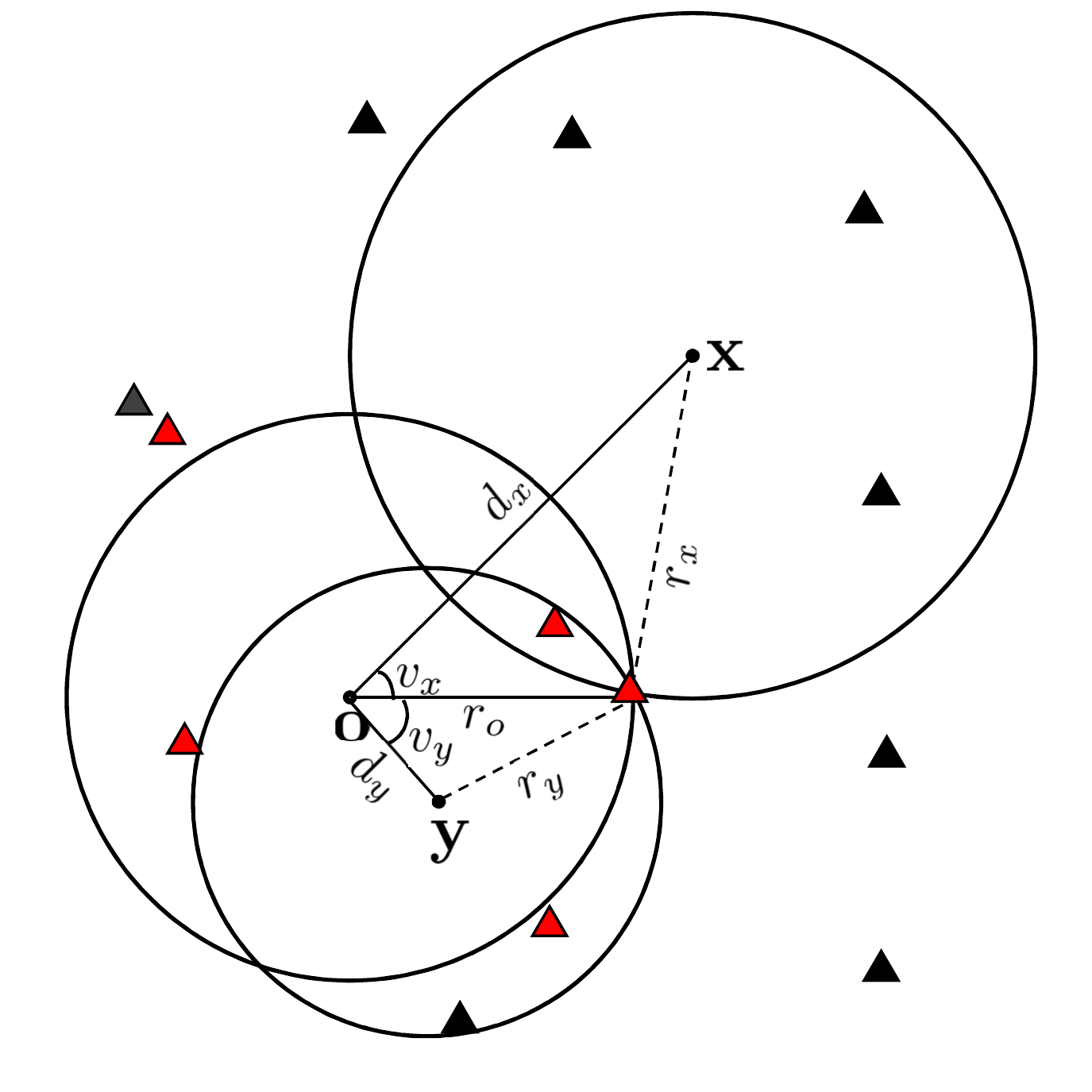}
\end{subfigure}%
\hspace{0.5cm}
\begin{subfigure}{0.3\textwidth}
  \centering
  \includegraphics[width=\linewidth]{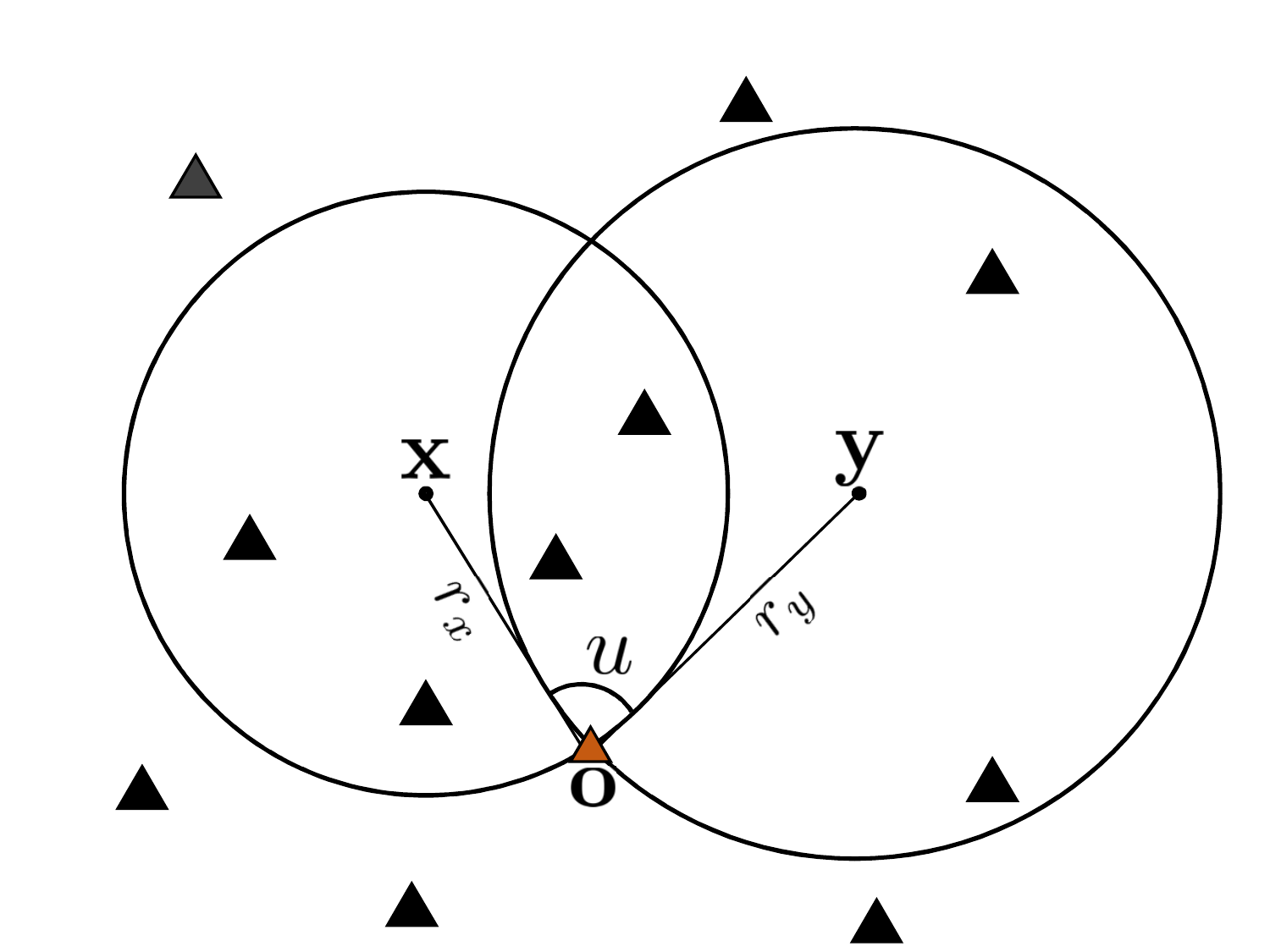}
\end{subfigure}%
\caption{\small (Left and center figures  are for the proof in Appendix~\ref{app:LoadTagRRH}.) The typical user is located at $\nbo$, while $\nbx, \nby \in \fiu$ are random user locations. The red triangles represent the serving AP locations of the typical user. The illustration is for the third nearest serving AP for the typical user. (Right figure is for the proof in Appendix~\ref{app:LoadTypRRH}.)  $\nbx$ and $\nby$ correspond to user locations and $\nbo$ correspond to the typical AP location. The number of APs in each non-overlapping region follow independent Poisson distribution.}
\label{fig:TagAPLoad_M1}
\end{figure*}

The mean load of the $N$-th closest serving AP to the typical user at $\nbo$ is given as 
\begin{align*}
\nbbE[K_N] & =  \nbbE\left[\sum_{\nbx \in \fiu} \mathbf{1}\left(|\fir \cap \ncalB_{r_x}(\nbx)| \leq \Ns-1\right) \mathbf{1}\left(|\fir \cap \ncalB_{r_o}(\nbo)|= N-1\right) \right] \\
& = \nbbE_{\fiu}\left[\sum_{\nbx \in \fiu} \nbbE_{\fir}\left[\mathbf{1}\left(|\fir \cap \ncalB_{r_x}(\nbx)| \leq \Ns-1\right) \mathbf{1}\left(|\fir \cap \ncalB_{r_o}(\nbo)|= N-1\right) \right]\right],
\end{align*}
where $r_o = \|\nbr - \nbo\|$, $r_x = \|\nbr - \nbx\| = \sqrt{r_o^2 + d_x^2 - 2 r_o d_x \cos(v_x)}$, $d_x = \|\nbx - \nbo\|$ (please refer to Fig.~\ref{fig:TagAPLoad_M1} (left)).
Conditioned on the location of the AP at $\nbr$, we expand the inner expectation as 
\begin{align*}
& \nbbE_{\{\fir \setminus \nbr\}}\left[\sum_{n=0}^{N-1} \mathbf{1}_{\left(|\fir \cap \ncalB_{r_o}(\nbo) \cap \ncalB_{r_x}(\nbx)|= n\right)} \mathbf{1}_{\left(|\fir \cap \{\ncalB_{r_o}(\nbo) \setminus \ncalB_{r_x}(\nbx)\}|= N-n-1\right)} \mathbf{1}_{\left(|\fir \cap \{\ncalB_{r_x}(\nbx) \setminus \ncalB_{r_o}(\nbo)\}| \leq \Ns-n-1\right)}\right] \\
= & \sum_{n=0}^{N-1} \frac{(\lamr |\ncalB_{r_o}(\nbo) \cap \ncalB_{r_x}(\nbx)|)^n}{n!} e^{-\lamr |\ncalB_{r_o}(\nbo) \cap \ncalB_{r_x}(\nbx)|} \frac{(\lamr |\ncalB_{r_o}(\nbo) \setminus \ncalB_{r_x}(\nbx)|)^{N-n-1}}{n!} e^{-\lamr |\ncalB_{r_o}(\nbo) \setminus \ncalB_{r_x}(\nbx)|} \\
& \sum_{l=0}^{\Ns-n-1}  \frac{(\lamr |\ncalB_{r_x}(\nbx) \setminus \ncalB_{r_o}(\nbo)|)^{\Ns-n-1}}{l!} e^{-\lamr |\ncalB_{r_x}(\nbx) \setminus \ncalB_{r_o}(\nbo)|} \\
= & \sum_{n=0}^{N-1} \ppmf(n, \lamr{\tt AoI}_2(r_o, d_x, v_x)) \ppmf(N-n-1, \lamr(\pi r_o^2 - {\tt AoI}_2(r_o, d_x, v_x))) \\
& \pcmf(\Ns-n-1, \lamr(\pi r_x^2 - {\tt AoI}_2(r_o, d_x, v_x))) = h_{\tt tag, m_1}(r_o = \|\nbr\|, d_x, v_x, N), \numberthis
\label{eq:TagM1Decomp}
\end{align*}
where the second step follows from the fact that $\{\fir \setminus \nbr\}$ is a homogeneous PPP with density $\lamr$ and the regions in each indicator function are non-overlapping. 
The final result follows in two step: first we decondition over the point $\nbr$, then we decondition over $\fiu$. Formally, $\nbbE[K_N] =$ 
\begin{align*}
 & \nbbE_{\fiu} \left[\sum_{\nbx \in \fiu} \int_{\nbr \in \R^2} h_{\tt tag, m_1}(\|\nbr\|, d_x, v_x, N) \lamr {\rm d}\nbr \right] = \nbbE_{\fiu} \left[\sum_{\nbx \in \fiu} 2 \pi \lamr \int_{r_o=0}^{\infty} h_{\tt tag, m_1}(r_o, d_x, v_x, N) r_o{\rm d}r_o \right] \\
= &  2 \pi \lamu \lamr \int_{r_o = 0}^{\infty} {\rm d}r_o \int_{d_x = 0}^{\infty} {\rm d}d_x \int_{v_x = 0}^{2 \pi} {\rm d}v_x h_{\tt tag, m_1}(r_o, d_x, v_x, N)r_o d_x,
\end{align*}
where the last step follows from the application of Campbell's theorem.

The second moment of the load of the $N$-th nearest serving AP is $\nbbE[|K_N|^2] =$
{\small
\begin{align*}
 & \nbbE\bigg[\bigg|\sum_{\nbx \in \fiu} \mathbf{1}_{ |\fir \cap \ncalB_{r_x}(\nbx)| \leq \Ns-1} \mathbf{1}_{ |\fir \cap \ncalB_{r_o}(\nbo)|= N-1}\bigg|^2\bigg] = \nbbE\bigg[\mathop{\sum_{\nbx \in \fiu}}_{\nby \in \fiu} \mathbf{1}_{ |\fir \cap \ncalB_{r_x}(\nbx)| \leq \Ns-1 } \mathbf{1}_{ |\fir \cap \ncalB_{r_y}(\nby)| \leq \Ns-1 } \mathbf{1}_{ |\fir \cap \ncalB_{r_o}(\nbo)|= N-1 }\bigg] \\
= & \nbbE\bigg[\mathop{\sum_{\nbx \in \fiu}^{\nbx \neq \nby}}_{\nby \in \fiu} \mathbf{1}_{ |\fir \cap \ncalB_{r_x}(\nbx)| \leq \Ns-1 } \mathbf{1}_{ |\fir \cap \ncalB_{r_y}(\nby)| \leq \Ns-1 } \mathbf{1}_{ |\fir \cap \ncalB_{r_o}(\nbo)|= N-1 }\bigg]  + \nbbE\bigg[ \sum_{\nbx \in \fiu} \mathbf{1}_{ |\fir \cap \ncalB_{r_x}(\nbx)| \leq \Ns-1} \mathbf{1}_{ |\fir \cap \ncalB_{r_o}(\nbo)|= N-1} \bigg].
\end{align*}
}
On the right hand side of the above equation, the second summation term is the mean $\nbbE[K_N]$, which has been derived above. 
We focus on deriving an expression for the first term (please refer to Fig.~\ref{fig:TagAPLoad_M1} (center)). Using \eqref{eq:Ang_2Circles}, let us define $u_x = u(r_o, d_x, v_x), u_y = u(r_o, d_y, v_y)$, and 
\begin{equation}
u_{xy} =
\begin{dcases}
u_x + u_y, & \{v_x < \pi, v_y > \pi\},  \\
u_x + u_y, & \{v_x > \pi, v_y < \pi\}, \\
|u_x - u_y|, & \text{Otherwise}.
\end{dcases}
\end{equation}
Further, let us denote the region of intersection of three circles as ${\tt RoI_{oxy}} = \{\ncalB_{r_o}(\nbo) \cap  \ncalB_{r_x}(\nbx) \cap \ncalB_{r_y}(\nby)\}$, the region exclusive to both the circles centered at $\nbo$ and $\nbx$ as ${\tt RoI_{ox}} = \{\ncalB_{r_o}(\nbo) \cap  \ncalB_{r_x}(\nbx)\} \setminus {\tt RoI_{oxy}}$, the regions exclusive to both the circles at  $\nbo$ and $\nby$ as ${\tt RoI_{oy}}= \{\ncalB_{r_o}(\nbo) \cap  \ncalB_{r_y}(\nby)\} \setminus {\tt RoI_{oxy}}$, and the common region exclusive to circles at $\nbx$ and $\nby$ as ${\tt RoI_{xy}}= \{\ncalB_{r_x}(\nbx) \cap  \ncalB_{r_y}(\nby)\} \setminus {\tt RoI_{oxy}}$.
Conditioned on $\fiu$ and $\nbr$, we write 
\begin{align*}
& \nbbE_{\{\fir \setminus \nbr\}}\left[\mathbf{1}_{ |\fir \cap \ncalB_{r_x}(\nbx)| \leq \Ns-1 } \mathbf{1}_{ |\fir \cap \ncalB_{r_y}(\nby)| \leq \Ns-1 } \mathbf{1}_{ |\fir \cap \ncalB_{r_o}(\nbo)|= N-1 }\right] \\
\stackrel{(b)}{=} & \nbbE_{\{\fir \setminus \nbr\}}\bigg[\sum_{n=0}^{N-1} \mathbf{1}_{ |\fir \cap  {\tt RoI_{oxy}}| = n} \!\!\!\! \sum_{m=0}^{N-n-1} \!\!  \mathbf{1}_{|\fir \cap {\tt RoI_{ox}}| = m}  \!\!\!\! \sum_{p=0}^{N-m-n-1} \!\!\!\!  \mathbf{1}_{|\fir \cap {\tt RoI_{oy}}| = p} \mathbf{1}_{|\fir \cap  \{\ncalB_{r_o}(\nbo) \setminus  \{\ncalB_{r_x}(\nbx) \cup \ncalB_{r_y}(\nby)\} \} \}| = N-n-m-p-1} \\
& \quad \quad \sum_{q=0}^{\min\{\Ns-n-m-1, \Ns-n-p-1\}} \mathbf{1}_{|\fir \cap {\tt RoI_{xy}} | = q} \mathbf{1}_{|\fir \cap  \{\ncalB_{r_x}(\nbx) \setminus  \{\ncalB_{r_o}(\nbo) \cup \ncalB_{r_y}(\nby)\} \} \}| \leq \Ns-n-m-q-1} \\
& \quad \quad  \mathbf{1}_{|\fir \cap  \{\ncalB_{r_y}(\nby) \setminus  \{\ncalB_{r_o}(\nbo) \cup \ncalB_{r_x}(\nbx)\} \} \}| \leq \Ns-n-p-q-1}
\bigg] \\
\stackrel{(c)}{=} & \sum_{n=0}^{N-1} \ppmf(n, \lamr {\tt AoI}_3(r_o, r_x, r_y, v_x, v_y))  \\
& \times \sum_{m=0}^{N-n-1} \ppmf(m, \lamr ({\tt AoI}_2(r_o, r_x, v_x) - {\tt AoI}_3(r_o, r_x, r_y, v_x, v_y))) \\
& \times \sum_{p=0}^{N-n-m-1} \ppmf(p, \lamr ({\tt AoI}_2(r_o, r_y, v_y) - {\tt AoI}_3(r_o, r_x, r_y, v_x, v_y))) \\
& \times \ppmf(N-n-m-p-1, \lamr (\pi r_o^2 - {\tt AoI}_2(r_o, r_x, v_x) - {\tt AoI}_2(r_o, r_y, v_y) + {\tt AoI}_3(r_o, r_x, r_y, v_x, v_y))) \\
&  \times \sum_{q=0}^{\min\big\{\substack{\Ns-n-m-1,\\ \Ns-n-p-1}\big\}} \ppmf(q ,\lamr  ({\tt AoI}_2(r_x, r_y, u_{xy}) - {\tt AoI}_3(r_o, r_x, r_y, v_x, v_y))) \\
&  \times \ppmf(\Ns-n-m-q-1, \lamr (\pi r_x^2 - {\tt AoI}_2(r_o, r_x, v_x) - {\tt AoI}_2(r_x, r_y, u_{xy}) + {\tt AoI}_3(r_o, r_x, r_y, v_x, v_y))) \\
&  \times \ppmf(\Ns-n-p-q-1, \lamr (\pi r_y^2 - {\tt AoI}_2(r_o, r_y, v_y) - {\tt AoI}_2(r_x, r_y, u_{xy}) + {\tt AoI}_3(r_o, r_x, r_y, v_x, v_y)))\\
& = h_{\tt tag, m_2}(r_o, d_x, d_y, v_x, v_y, N), \numberthis
\label{eq:htagM2}
\end{align*}
where $r_i = \sqrt{d_i^2 + r_o^2 - 2 r_o d_i \cos(v_i)}$ for $i \in \{x, y\}$, the function ${\tt AoI}_2\left(\cdot, \cdot, \cdot\right)$ is given in \eqref{eq:AoI_2Circles}, the area of intersection of three circles is evaluated as per the procedure presented in Appendix~\ref{app:AreaOfInt3}.
The step $(a)$ follows from the fact that the regions in indicator functions are non-overlapping and $\fir \setminus \nbr$ is a homogeneous PPP with density $\lamr$.
Similar to the derivation of the first moment, we obtain the final expression for the second moment by deconditioning over $\nbr$ and then over $\fiu$ by applying Campbell's theorem.

\subsection{Proof of Lemma~\ref{lem:LoadTypRRH}}\label{app:LoadTypRRH}
Much of the derivation can be done on the similar lines as that of Appendix~\ref{app:LoadTagRRH}. Since $\fir$ is a homogeneous PPP, it is translation invariant. Hence, we assume that the typical AP is located at the origin. The mean load of the AP is $\nbbE[K_o] = \nbbE\left[\sum_{\nbx \in \fiu} \nbbE\left[\mathbf{1}\left(|\fir \cap \ncalB_{\|\nbx\|} (\nbx)| \leq \Ns-1 \right) \right]\right] =$ 
\begin{align*}
 2 \pi \lamu \int_{r=0}^{\infty} \sum_{l=0}^{\Ns-1} \frac{(\pi \lamr r^2 )^{l}}{l!} \exp(-\pi \lamr r^2) r{\rm d}r   \stackrel{(a)}{=} \frac{\lamu}{\lamr} \sum_{l=0}^{\Ns-1} \int_{u=0}^{\infty} \frac{u^l}{l!} \exp(-u) {\rm d}u =  \frac{\Ns \lamu}{\lamr},
\end{align*}
where $(a)$ follows from replacing $u = \pi \lamr r^2$.
The second moment of the load can be written as 
\begin{align*}
\nbbE[K_o^2] = &  \nbbE\bigg[\bigg|\sum_{\nbx \in \fiu} \mathbf{1}\left(|\fir \cap \ncalB_{\|\nbx\|} (\nbx)| \leq \Ns-1 \right)\bigg|^2\bigg] \\
= &  \nbbE\bigg[\mathop{\sum_{\nbx \in \fiu}^{\nbx \neq \nby}}_{{\nby \in \fiu}} \nbbE_{\fir}\left[\mathbf{1}_{|\fir \cap \ncalB_{\|\nbx\|} (\nbx)| \leq \Ns-1} \mathbf{1}_{|\fir \cap \ncalB_{\|\nby\|} (\nby)| \leq \Ns-1}\right]\bigg] + \underbrace{\nbbE\bigg[\sum_{\nbx \in \fiu} \nbbE\left[\mathbf{1}_{|\fir \cap \ncalB_{\|\nbx\|} (\nbx)| \leq \Ns-1} \right]\bigg]}_{\nbbE[K_o]}.
\end{align*}
We can decompose the inner expectation $\nbbE_{\fir}\left[\mathbf{1}_{|\fir \cap \ncalB_{\|\nbx\|} (\nbx)| \leq \Ns-1} \mathbf{1}_{|\fir \cap \ncalB_{\|\nby\|} (\nby)| \leq \Ns-1}\right] = $
\begin{align*}
&\sum_{l=0}^{\Ns-1}\nbbE_{\fir}\bigg[\mathbf{1}_{|\fir \cap \{\ncalB_{\|\nbx\|} (\nbx) \cap \ncalB_{\|\nby\|} (\nby)\}| = l} \mathbf{1}_{|\fir \cap \{ \ncalB_{\|\nby\|}(\nby) \setminus \{\ncalB_{\|\nbx\|} (\nbx) \cap \ncalB_{\|\nby\|} (\nby)\} | \leq \Ns-l-1}  \\
& \times  \mathbf{1}_{|\fir \cap \{ \ncalB_{\|\nbx\|}(\nbx) \setminus \{\ncalB_{\|\nbx\|} (\nbx) \cap \ncalB_{\|\nby\|} (\nby)\} | \leq \Ns-l-1} \bigg].
\end{align*}
From the above expression, we obtain the expression for $h_{\tt typ, m2}(r_x, r_y, u)$ by taking the expectation with respect to $\fir$.
Now, with the application of Campbell's formula, we get the final expression of the lemma as
\begin{align*}
\nbbE\left[\sum_{\nbx \in \fiu} \sum_{\nby \in \fiu}^{\nbx \neq \nby} h_{\tt typ, m2}(r_x, r_y, u)\right] = 2 \pi \lamu^2 \int_{r_x=0}^{\infty} \int_{r_y=0}^{\infty} \int_{u=0}^{2 \pi} h_{\tt typ, m2}(r_x, r_y, u)  {\rm d}u  r_y{\rm d}r_y r_x{\rm d}r_x,
\end{align*}
where $r_x, r_y,$ and $u$ are as depicted in Fig.~\ref{fig:TagAPLoad_M1} (Right).
\vspace{-0.6cm}

\subsection{Area of Intersection of Three Circles}\label{app:AreaOfInt3}
\begin{figure*}[!htb]
\centering
\begin{subfigure}{0.27\textwidth}
  \centering
  \includegraphics[width=\linewidth]{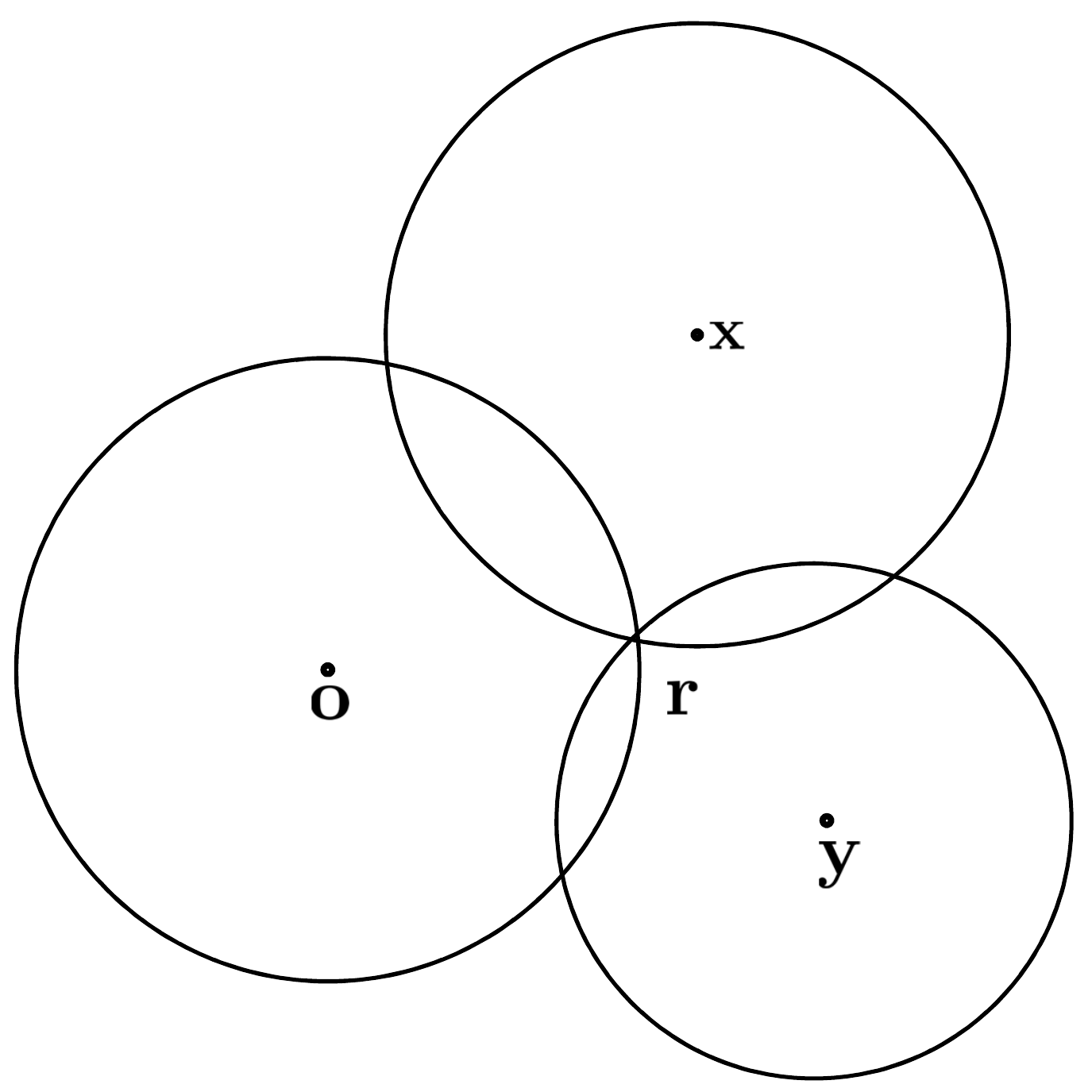}
\end{subfigure}
\hspace{0.1cm}
\begin{subfigure}{0.27\textwidth}
  \centering
  \includegraphics[width=\linewidth]{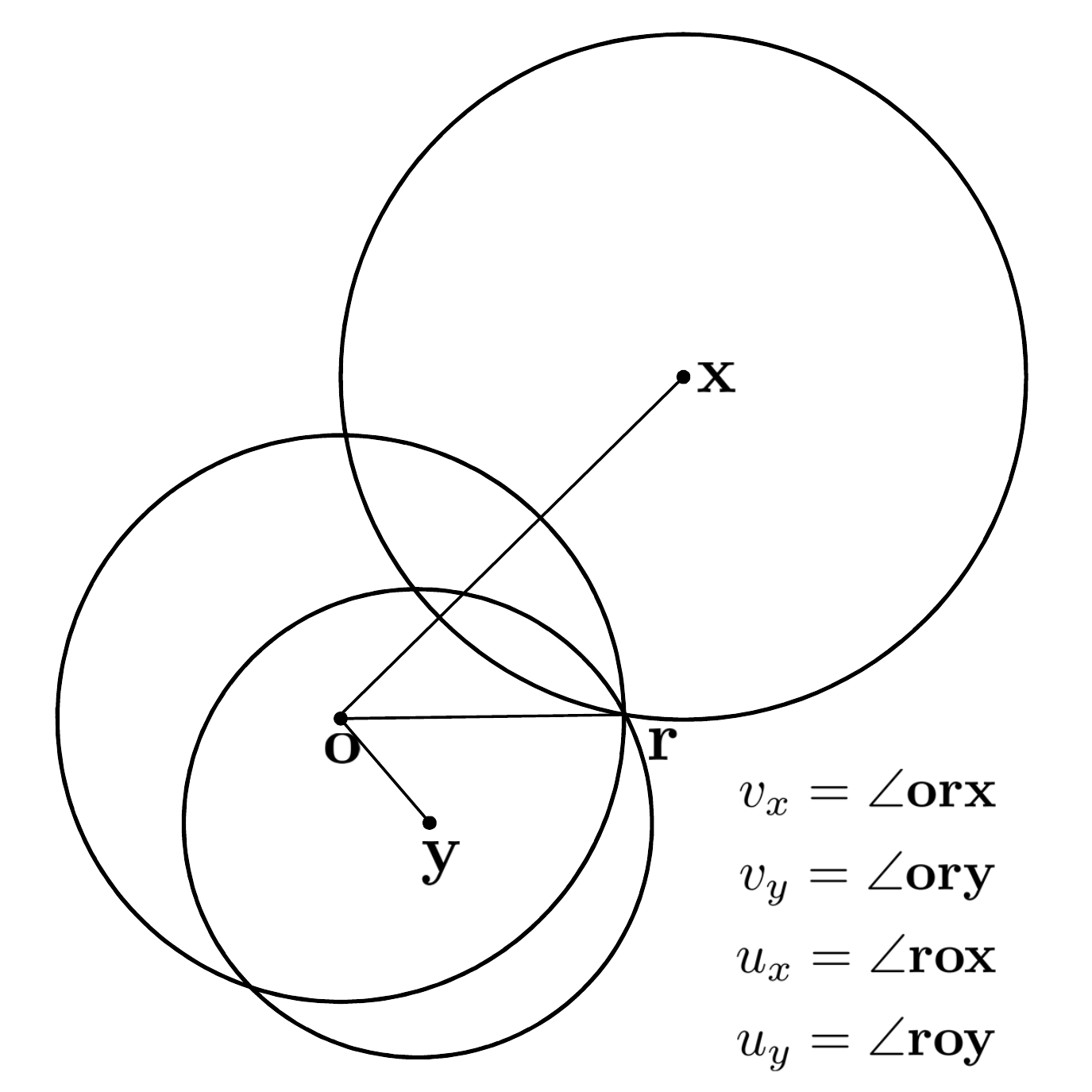}
\end{subfigure}%
\hspace{0.1cm}
\begin{subfigure}{0.27\textwidth}
  \centering
  \includegraphics[width=\linewidth]{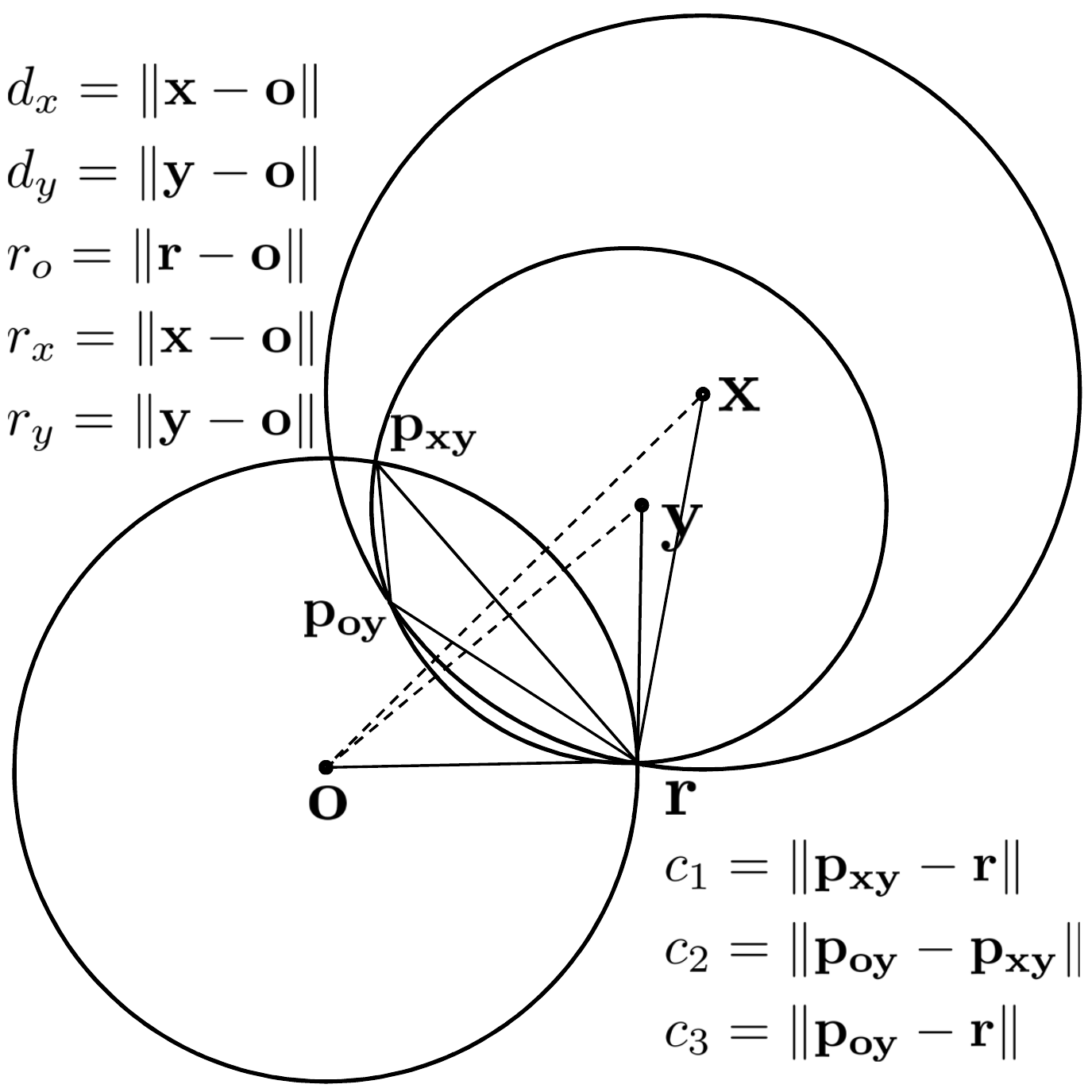}
\end{subfigure}%
\caption{\small Three possible configurations of the intersection of three circles with one common point of intersection $\nbr$. The fourth configuration in which the smallest circle lies inside the two bigger circles has been ignored as it is a zero probability event. All relevant distances and angles have been marked in the figure.}
\label{fig:CirInt}
\end{figure*}

Due to the constraint that the three circles have a common point of intersection, the common area of intersection will be either of the following three cases: (1) a point with area zero, (2) a lens, or (3) a circular triangle. All three cases are presented in Fig.~\ref{fig:CirInt}.
When the common area of intersection is a circular triangle (right most case in Fig.~\ref{fig:CirInt}), the area is given as~\cite{fewell2006area}
\begin{align*}
{\tt AoI}_3(r_o, r_x, r_y, v_x, v_y)  = & \frac{1}{4} \sqrt{(c_1 + c_2 + c_3) (c_2 + c_3 - c_1) (c_1 + c_3 - c_2) (c_1 + c_2 - c_3)} \\ 
& + r_o^2 \arcsin\frac{c_1}{2r_o} - \frac{c_1}{4} \sqrt{4 r_o^2 - c_1^2} + r_y^2 \arcsin\frac{c_2}{2r_y} - \frac{c_2}{4} \sqrt{4 r_y^2 - c_2^2} \\
& + r_x^2 \arcsin\frac{c_3}{2r_x} - \frac{c_3}{4} \sqrt{4 r_x^2 - c_3^2},
\end{align*}
where $c_1, c_2, c_3$ are chord lengths as denoted in the figure. Please note that the first two cases are special cases of the third case, e.g. we can get the second case by replacing $c_2 = 0$ and $c_1 = c3$. Similarly, in the first case, $c_1 = c_2 = c_3 = 0$. Further, $c_i$s are functions of $r_o, r_x, r_y, v_x, v_y$. The procedure for determining them is outlined in~\cite{fewell2006area} that is followed in this work.

\subsection{Proof of Proposition~\ref{prop:RateCov}}\label{app:RateCov}
Ignoring the pilot contamination term in the expression of achievable rate in \eqref{eq:RoLB}, we can write the rate coverage as 
\begin{align*}
\Rcinf = \nbbP\bigg[ &\frac{\rd \Na}{2^{\Tr}-1} \left(\sum\limits_{\nbr_l \in \Phi_{ro}}  \sqrt{\gamma_{lo} (1 - 2^{-\Cf/k_l})/{K_{\tt max}}}\right)^2 - \rd \Na\sum\limits_{\nbr_l \in \Phi_{ro}} {\gamma_{lo}} 2^{-\Cf/k_l}/{K_{\tt max}}  \\
& - \rd \sum\limits_{\nbr_l \in \Phi_{ro}} \beta_{lo}  -1 \geq \rd \sum\limits_{\nbr_l \in \Phi_{ro}^C} \beta_{lo}  \bigg].\numberthis
\label{eq:app:RateCov}
\end{align*}
To proceed further, we first condition on the distance to the $\Ns$-th serving AP $d_{o\Ns}$.
Conditioned on this distance, we replace $\sum\limits_{\nbr_l \in \Phi_{ro}^C} \beta_{lo}$ by its mean which is given as 
$
\nbbE_{\Phi_{ro}^C}[\sum\limits_{\nbr_l \in \Phi_{ro}^C} \beta_{lo}] = 2 \pi \lamr \int_{r = d_{o\Ns}}^{\infty} l(r) r{\rm d}r.
$
This result follows from the application of Campbell's theorem.
Note that using the mean instead of the exact expectation has marginal impact on the accuracy of the result as $\sum\limits_{\nbr_l \in \fir} \beta_{lo}$ is dominated by contributions from the nearest $\Ns$ serving APs. Hence, we write 
\begin{align*}
\sum\limits_{\nbr_l \in \fir} \beta_{lo} \approx \sum\limits_{\nbr_l \in \Phi_{ro}} \beta_{lo} + \nbbE_{\Phi_{ro}^C}[\sum\limits_{\nbr_l \in \Phi_{ro}^C} \beta_{lo}].
\end{align*}
Next, the loads for the different serving APs are correlated. Hence, to get the exact result, we need to evaluate \eqref{eq:app:RateCov} with respect to the joint distribution of $\{K_i\}_{i=1}^{\Ns}$. However, obtaining this joint distribution is not tractable. 
Hence, we exactly consider the load of the nearest AP and replace the load of the rest of the APs by its effective mean. For the $i$-th nearest AP, the effective mean is given as 
$
\bar{K}_i = 1 +  \sum_{k_i = 0}^{\infty} \min\{k_i, K_{\tt max}\} \nbbP[K_i = k_i],
$
where $K_i$ follows negative binomial distribution whose $\pmf$ is determined using the moment matching method presented in Sec.~\ref{subsec:TagAP}.
Under the above two approximations, conditioned on the distances to the serving APs and the load of the nearest AP to the typical user, the rate coverage is given as 
\begin{align*}
\Rcinf = \nbbE_{k_1, d_{o1}, \ldots, d_{o\Ns}}\bigg[\mathbf{1}\bigg(2 \pi \lamr \int\limits_{d_{o\Ns}}^{\infty} l(r) r{\rm d}r \leq h_{\tt cov}(k_1, d_{o1}, d_{o2}, \ldots, d_{o\Ns}\bigg)\bigg], \numberthis
\label{eq:app:RateCov2}
\end{align*}
where $ h_{\tt cov}(k_1, d_{o1}, d_{o2}, \ldots, d_{o\Ns})$ is given by \eqref{eq:RHSofRateCov}.
Note that conditioned on $d_{o\Ns}$, $d_{oi}$ for $1 \leq i \leq \Ns -1$ are i.i.d. distributed with the following ${\tt PDF}$~\cite{Haenggi2013}
\begin{align*}
f_{D_{oi}}(d_{oi}) = \frac{2 d_{oi}}{d_{o\Ns}^2}, \quad 0 \leq d_{oi} \leq d_{o\Ns}.
\end{align*}
Further, the ${\tt PDF}$ of $D_{o\Ns}$ is given as~\cite{Haenggi2013}
\begin{align*}
f_{D_{o\Ns}}(d_{o\Ns}) = \frac{2}{\Gamma(\Ns)} (\pi \lamr)^{\Ns} d_{o\Ns}^{2\Ns -1} \exp(-\pi \lamr d_{o\Ns}^2).
\end{align*}
We evaluate the expectation in \eqref{eq:app:RateCov2} using the aforementioned distance distributions along with the $\pmf$ of the load $K_1$ associated with the nearest tagged AP.
\bibliographystyle{IEEEtran}
\bibliography{DiMiMO}

\end{document}